\documentclass{article}
\usepackage{amsmath,amssymb}
\usepackage{color}
\usepackage{cite}
\usepackage{hyperref}

\usepackage{amsthm}
\theoremstyle{definition}
\newtheorem{thm}{Theorem}
\newtheorem{lmm}{Lemma}
\newtheorem{dfn}{Definition}

\makeatletter
\@addtoreset{equation}{section}
\renewcommand{\theequation}{\thesection.\@arabic\c@equation}
\makeatother
\makeatletter
\renewcommand\appendix{\par%\newpage
  \setcounter{section}{0}%
  \setcounter{subsection}{0}%
  \gdef\thesection{Appendix \@Alph\c@section }
  \renewcommand{\theequation}
  {\Alph{section}.\arabic{equation}}
}
\makeatother
\def \be {\begin{equation}}
\def \ee {\end{equation}}
\def \ba {\begin{array}}
\def \ea {\end{array}}
\def \bea{\begin{eqnarray}}
\def \eea{\end{eqnarray}}

\def \G {\Gamma}

\def \m {\mu}
\def \n {\nu}

\def \L {\Lambda}
\def \s {\sigma}

\def \r {\rho}
\def \o {\omega}
\def \O {\Omega}
\def \th {\theta}
\def \Th {\Theta}

\def \p {\partial}
\def \f {\frac}
\def \na {\nabla}

\def \nn {\nonumber}

\def \ma {\mathcal}

\def \lt {\left}
\def \rt {\right}

\def \sr {\sqrt}
\def \td {\tilde}
\def \hs {\hspace}
\def \pp {\propto}
\def \inf {\infty}

\title{\textbf{Note on Thermodynamics Method of Black Hole/CFT Correspondence}}
\author{
Bin Chen$^{1,2}$\footnote{bchen01@pku.edu.cn},\,
Zhao Xue$^{1}$\footnote{xuezhao2012@pku.edu.cn}\,
and
Jia-ju Zhang$^{1}$\footnote{jjzhang@pku.edu.cn}
}
\date{}

\begin{document}

\maketitle

\begin{center}
{\it
$^{1}$Department of Physics and State Key Laboratory of Nuclear Physics and Technology, Peking University, No.5 Yiheyuan Road, Beijing 100871, P.R. China\\
\vspace{2mm}
$^{2}$Center for High Energy Physics, Peking University, No.5 Yiheyuan Road, Beijing 100871, P.R. China
}
\vspace{10mm}
\end{center}

\begin{abstract}

In the paper we further refine the thermodynamics method of black hole/CFT correspondence. We show that one can derive the central charges of different holographic pictures directly from the entropy product $S_+S_-$ if it is mass-independent, for a black hole in the Einstein gravity or the gravity without diffeomorphism anomaly. For a general black hole in the Einstein gravity that admits holographic descriptions, we show that the thermodynamics method and asymptotic symmetry group (ASG) analysis can always give consistent results in the extreme limit.  Furthermore, we discuss the relation between black hole thermodynamics and the hidden conformal symmetry. We show that the condition $T_+A_+=T_-A_-$, with $A_\pm$ being the outer and inner horizon areas, is the necessary, but not sufficient, condition for a black hole to have the hidden conformal symmetry. In particular, for the Einstein(-Maxwell) gravity $T_+A_+=T_-A_-$ is just the condition $T_+S_+=T_-S_-$, with $S_\pm$ being the outer and inner horizon entropies, which is the condition for the entropy product $S_+S_-$ being mass-dependent. When there exists the hidden conformal symmetry in the low-frequency scattering off the generic non-extremal black hole, it always leads to the same temperatures of dual CFT as the ones got from the thermodynamics method.

\end{abstract}

\baselineskip 18pt
\thispagestyle{empty}

\newpage

%\tableofcontents

\section{Introduction}

One of great achievements of string theory is that the Bekenstein-Hawking entropy of some kinds of BPS, thus extremal, charged black holes could be counted microscopically in a two-dimensional (2D) holographic conformal field theory (CFT) dual \cite{Strominger:1996sh}. Such a microscopic counting in a 2D CFT was generalized to neutral rotating black holes, i.e. Kerr black hole, in the so-called Kerr/CFT correspondence  in the past few years \cite{Guica:2008mu,Castro:2010fd}. There have been many generalizations and extensions of the Kerr/CFT correspondence, and one could see more details in the reviews \cite{Bredberg:2011hp,Compere:2012jk} and references therein.

Different from the brane constructions in the study of the black hole in string theory, the setting up of the Kerr/CFT correspondence has nothing to do with string theory. The information on the holographic picture of black hole was read in different ways.
The central charges of dual CFT are usually obtained from doing asymptotic symmetry group (ASG) analysis \cite{Brown:1986nw,Brown:1986ed,Barnich:2001jy,Barnich:2007bf,Guica:2008mu,Carlip:2011ax,Carlip:2011vr} of the near horizon geometry of the extreme black hole.
%Note that the first derivation of central charges using ASG method was done for AdS$_3$ spacetime in \cite{Brown:1986nw,Brown:1986ed}, in the so-called Barnich-Brandt-Compere (BBC) formalism \cite{Barnich:2001jy,Barnich:2007bf} one has covariant systematic techniques of treating ASG, and there is also the stretched horizon formulism proposed in \cite{Carlip:2011ax,Carlip:2011vr}.
Strictly speaking, the central charges are the ones of dual CFT for the extremal black holes. It has been believed that if the form of the central charges can be written in terms of quantized charges, then it should still be true for the non-extremal black holes. This argument has gained support from recent study on the black hole/CFT correspondence via thermodynamics of both outer and inner horizons, which may give the central charges for generic cases.
To read the temperatures of dual CFT is even more tricky. For extremal black holes, one may use the Frolov-Thorne vacuum \cite{Frolov:1989jh,Guica:2008mu} to read them.  For generic non-extreme black holes, one can only read them from the hidden conformal symmetry \cite{Castro:2010fd,Chen:2010fr} in the low frequency scattering off the black holes. But for some cases like the black rings, the hidden conformal symmetry can not be found even though the dual pictures of the black objects are expected. It is fair to say that for generic non-extremal black holes, their holographic pictures were established not on a solid ground.

Very recently, the investigation of holographic pictures from thermodynamics point of view open a new window and shed more light on the black hole/CFT correspondence.\footnote{In \cite{daCunha:2010jj} there was trial of deriving the asymptotic boundary conditions in Kerr/CFT correspondence using the surface integral from the outer horizon thermodynamics.}
 It had been noted long ago that the inner horizon may play an important role in counting the entropy of the outer horizon
\cite{Cvetic:1996kv,Larsen:1997ge,Cvetic:1997uw,Cvetic:1997xv}. And it was suggested in \cite{Cvetic:2009jn,Cvetic:2010mn} the property that the entropy product $S_+S_-$ being mass-independent, with $S_\pm$ as the Bekenstein-Hawking entropies of the outer and inner horizons, has nontrivial implications for the holographic descriptions of black holes. The first law of thermodynamics of the inner horizon has been checked for five-dimensional black rings in \cite{Castro:2012av} and for some kinds of three-dimensional (3D) black holes in \cite{Detournay:2012ug}.
It was proved in \cite{Chen:2012mh} that under reasonable assumption the first law of thermodynamics of the outer horizon always indicates that of the inner horizon, and the mass-independence of the entropy product $S_+S_-$ is equivalent to the condition $T_+S_+=T_-S_-$, with $T_\pm$ being the Hawking temperatures of the outer and inner horizons. Based on the previous works
\cite{Cvetic:1996kv,Larsen:1997ge,Cvetic:1997uw,Cvetic:1997xv,Kastor:1997gt,Lu:2008jk,Hartman:2008pb,Azeyanagi:2008kb,
Chow:2008dp,Azeyanagi:2008dk,Cvetic:2009jn,Cvetic:2010mn}
it was propose in \cite{Chen:2012mh} that  for a Kerr black hole there is the first law of thermodynamics for the dual CFT. This first law could be obtained from the combinations of the first laws of the outer and inner horizons. It could finally be written in the following suggestive form
\be \label{x1}
dJ=T_L^J d S_L-T_R^J d S_R,
\ee
with $S_{L,R}=\f{1}{2}(S_+ \pm S_-)$ and $T_{L,R}^J$ being respectively the entropies and the temperatures of the left- and right-moving sectors of the CFT. It is amazing to see that the temperatures $T_{L,R}^J$ are exactly the same as the ones obtained from the hidden conformal symmetry.

This so-called thermodynamics method can be applied to the Reissner-Nordstr\"om (RN) black holes and other more complicated cases with multiple angular momenta and charges. For the black hole with $U(1)$ charges, one has to require the ``quantization" condition\footnote{There was a similar quantization condition in \cite{Azeyanagi:2008dk} where the Kerr/CFT correspondence was applied to extremal black holes in string theory.}
such that the first law of dual CFT could always be written as
\be
dN=T_L^N d S_L-T_R^N d S_R,
\ee
with $N$ being an integer-valued charge. This refinement resolves the puzzle on the ambiguity in deciding the CFT dual of RN-like black holes, and is consistent with the quantization rule on the angular momentum. This leads to the CFT duals of RN black holes in arbitrary dimensions and four-dimensional (4D) dyonic RN/CFT \cite{Chen:2012ps,Chen:2012pt}.

For the black holes with multiple holographic pictures, the effectiveness of the thermodynamics method is remarkable. For every conserved $U(1)$ charge, either an angular momentum or a $U(1)$ gauge charge, there could be an independent dual CFT picture\cite{Lu:2008jk,Hartman:2008pb,Krishnan:2010pv,Chen:2010ywa}. In order to read this picture, one just needs to turn off other charges in the thermodynamics law and find the information of dual CFT straightforwardly. Moreover, it was proposed in \cite{Lu:2008jk,Chen:2011wm,Chen:2011kt} that if there exit $n$ charges except the mass, besides the $n$ elementary CFT pictures there will be generic pictures generated by $SL(n,Z)$ transformations acting on the elementary pictures. The $SL(n,Z)$ group could be identified as a subgroup of T-duality group $O(n,n,Z)$\cite{Chen:2011wm}, but it could include the electromagnetic duality, i.e. S-duality, group $SL(2,Z)$ as well\cite{Chen:2012ps}.
In the thermodynamics method, these generic picture could be set up easily by acting $SL(n,Z)$ transformations on the charges in the first laws directly. The underlying physical picture is that with respect to different kinds of perturbations the black hole responds
and behaves differently and gives corresponding CFT pictures.

The thermodynamics method has been checked in \cite{Chen:2012mh} for many well-established black hole/CFT correspondences, including three-dimensional (3D) BTZ, four-dimensional (4D) Kerr-Newman and five-dimensional (5D) Myers-Perry black holes
\cite{Strominger:1997eq,Lu:2008jk,Hartman:2008pb,Krishnan:2010pv,Wang:2010qv,Chen:2010xu,Chen:2010ywa,Chen:2011wm,Chen:2011kt},
and has also been applied to the study of holographic descriptions of black rings \cite{Chen:2012mh,Chen:2012yd}. One interesting question is to understand the relation between the thermodynamics method and other conventional ways in establishing the black hole/CFT correspondence. In this paper, we tried to address this issue.
In Section~\ref{s2}, after a brief review of the thermodynamics method, we derive a simple formula to get the central charges of the dual CFT from the product of the entropies of two horizons. Then we find that for a general form of 4D black hole metric, the thermodynamics method gives the same central charges in the extreme limit as the ones from the ASG analysis. Furthermore, we show that the relation $T_+A_+=T_-A_-$, where $A_\pm$ and $T_\pm$ are the areas and temperatures of the outer and inner horizons respectively, is the necessary but not sufficient condition for a black hole to have the hidden conformal symmetry. This relation is equivalent to the mass-independence of the entropy product in the Einstein(-Maxwell) gravity. Following \cite{Chen:2012ps,Chen:2012pt}, we  show that there are general pictures characterized by vectors ${\bf v}=(v_1,\cdots,v_n)$, with $\{v_1,\cdots,v_n\}$ being $n$ coprime integers, could be got easily in the thermodynamics method by considering different kinds of perturbations. Furthermore we will show that the ${\bf v}$ pictures are in one-to-one correspondence to the $SL(n,Z)$ generated pictures.
In Section~\ref{s3}, we generalize our discussions to higher dimensions.
We end with some discussions in Section ~\ref{s4}.

\section{Four-dimensional Black Holes}\label{s2}

In this section we consider the holographic descriptions of 4D black holes using the thermodynamics method. Though our treatment could be applied to many kinds of black holes, say with multiple $U(1)$ charges, we mainly focus on the rotating black hole with only one $U(1)$ charge.
Here we do not need the explicit form of 4D Kerr-Newman black hole and work with the metric of a general form, so our discussions could be generalized to other cases easily.

\subsection{General black hole}

We describe the general properties of 4D rotating charged black hole in Einstein gravity that will be used later. We start from the action of Einstein-Maxwell theory
\be
I=\f{1}{16\pi G_4} \int d^4x\sr{-g}(R-2\L)-\f{1}{16\pi} \int d^4x\sr{-g}F_{\m\n}F^{\m\n},
\ee
with $c=\hbar=1$, $G_4=\ell_p^2$. We include the cosmology constant $\L$ for generality. Suppose that the action has stationary black hole solution whose metric could be written in the Arnowitt-Deser-Misner (ADM) form
\be \label{ds4}
ds^2_4=-N^2 dt^2+h_{rr}dr^2+h_{\th\th}d\th^2+h_{\phi\phi} \lt( d\phi+N^\phi dt \rt)^2,
\ee
and we assume the black hole has $\p_t$ and $\p_\phi$ isometries. The CFT duals for a black hole with both electric and magnetic charges are subtle and have been considered in \cite{Chen:2012ps,Chen:2012pt}, so here we consider the black hole with only an electric charge. We write the electromagnetic potential as
\be \label{e60} A=A_t dt+A_\phi d\phi, \ee
with $\p_t A_\m=\p_\phi A_\m=0$. We suppose that the black hole has outer and inner horizons located at $r_\pm$, so that
\be \label{e62}
N^2=(r-r_+)(r-r_-)f_1^2, ~~~ h_{rr}=\f{f_2^2}{(r-r_+)(r-r_-)},
\ee
with $f_1,f_2$ being positive functions of $r,\th$, and being neither vanishing nor divergent in the region $r\geq r_-$.

The Hawking temperatures, the angular velocities, and the electric potentials of the outer and inner horizons are respectively
\bea \label{ht}
&& T_\pm=\f{r_+-r_-}{4\pi} \lt. \f{f_1}{f_2} \rt|_{r=r_\pm}, \nn\\
&& \O_\pm=-N^\phi |_{r=r_\pm},  \nn\\
&&\Phi_\pm=-A_\mu \xi^\mu|_{r=r_\pm},
\eea
with $\xi=\p_t-N^\phi \p\phi$. Here we suppose that $N^\phi|_{r\to\infty}=A_\mu \xi^\mu|_{r\to\infty}=0$, which are actually true only for asymptotically flat spacetime. For the Kerr-Newman-AdS(-dS) black holes, the definitions of angular velocities and the electric potential need minor modification, and will be discussed in Section~\ref{s4}. According to the zeroth law of thermodynamics of the black hole \cite{Bardeen:1973gs}, we must have $T_\pm,\O_\pm,\Phi_\pm$ everywhere the same at the horizons, i.e.\! they do not depend on the polar angle $\th$. In the Einstein gravity we have the entropies at the outer and inner horizons  being
\be \label{spm}
S_\pm=\f{A_\pm}{4 G_4}=\f{1}{4 \ell_p^2} \int d\th d\phi f_5 f_6 |_{r=r_\pm},
\ee
with $f_5= \sr{h_{\th\th}}, f_6= \sr{h_{\phi\phi}}$ being regular positive functions of $r,\th$ like $f_{1,2}$ defined before. Also, there are the mass $M$, the angular momentum $J$, and the electric charge $Q$ of the black hole, whose explicit form are not important to us. The essential point is that there are first laws of thermodynamics of the black hole at the outer and inner horizons written as
\bea \label{e1}
&&d M=T_+ d S_+ + \O_+d J+\Phi_+ d Q  \nn\\
&&\phantom{d M}=-T_- d S_- + \O_-d J+\Phi_- d Q.
\eea
It was proved that under reasonable assumptions the first law of the outer horizon always indicates that of the inner horizon \cite{Chen:2012mh}.

\subsection{Thermodynamics method}

The thermodynamics method of black hole/CFT duality was developed in \cite{Chen:2012mh,Chen:2012ps}, and here we summarize the essential features of the method and develop the method further.

According to \cite{Chen:2012ps}, in the thermodynamics method we have to use the integer charges, and so we rewrite the first laws (\ref{e1}) as
\bea \label{e9}
&&d M=T_+ d S_+ + \O_+^1 d N_1 + \O_+^2 d N_2  \nn\\
&&\phantom{d M}=-T_- d S_- + \O_-^1 d N_1 + \O_-^2 d N_2.
\eea
Here we define $N_2=Q/e$ to be an integer, with $e$ being the unit charge of the Maxwell theory. Correspondingly we have  made the definitions $\O_\pm^1=\O_\pm$, $N_1=J$, $\O_\pm^2=e\Phi_\pm$ for convenience. One can show easily that if $T_+S_+=T_-S_-$ is satisfied, then the entropy product $S_+S_-$ is mass-independent. Under the condition that $T_+S_+=T_-S_-$ is satisfied we define the entropy product function
\be \label{e3}
\ma F \equiv \f{S_+ S_-}{4\pi^2},
\ee
with $\ma F$ being a function of the charges $N_{1,2}$.

We firstly make the definitions \cite{Cvetic:1997uw,Cvetic:1997xv,Cvetic:2009jn}
\bea
&& T_{R,L}=\f{T_-T_+}{T_- \pm T_+},  \nn\\
&& S_{R,L}=\f{1}{2}(S_+ \mp S_-),  \nn\\
&& \O_{R}^{1,2}=\f{T_- \O_+^{1,2} + T_+ \O_-^{1,2}}{2(T_- + T_+)},  \nn\\
&& \O_{L}^{1,2}=\f{T_- \O_+^{1,2} - T_+ \O_-^{1,2}}{2(T_- - T_+)},
\eea
and rewrite the first laws of the outer and inner horizons as the right- and left-moving sectors
\bea \label{e8}
&&\f{1}{2}d M=T_R d S_R+\O_R^1 d N_1  +\O_R^2 d N_2 \nn\\
&&\phantom{\f{1}{2}d M}=T_L d S_L  +\O_L^1 d N_1  +\O_L^2 d N_2.
\eea

Keeping $N_2$ invariant, from the above first laws we could get
\be \label{e2}
d N_1=\f{T_L}{\O_R^1-\O_L^1}d S_L  - \f{T_R}{\O_R^1-\O_L^1}d S_R.
\ee
Up to now we have done nothing but some mathematical redefinitions. Then there are two nontrivial claims, which have no direct justifications.
\begin{itemize}
  \item $S_{R,L}=\f{1}{2}(S_+ \mp S_-)$ are the entropies of right- and left-moving sectors of CFT dual to the black hole \cite{Cvetic:1996kv,Larsen:1997ge,Cvetic:1997uw,Cvetic:1997xv,Cvetic:2009jn};
  \item Keeping $N_2$ invariant, in $N_1$ picture CFT there is the first law of thermodynamics \cite{Chen:2012mh,Chen:2012ps}
  \be \label{e4}
   d N_1=T_L^1 d S_L-T_R^1 d S_R,
  \ee
  with $T_{R,L}^1$ being right- and left-moving temperatures.
\end{itemize}
These are two basic assumptions of the thermodynamics method, which has no direct derivations now. Admitting the above two claims, we could derive the following result using (\ref{e2}).
\begin{itemize}
\item $T_{R,L}=\f{T_-T_+}{T_- \pm T_+}$ are proportional to the CFT temperatures $T_{R,L}^1$ with a common scale factor $R_1$ \cite{Cvetic:1997uw,Cvetic:1997xv,Cvetic:2009jn}, and explicitly we have
    \bea \label{e16}
    && R_1=\f{1}{\O_R^1-\O_L^1}=\f{T_-^2-T_+^2}{T_- T_+(\O_-^1 - \O_+^1)},   \nn\\
    && T_{R,L}^1=R_1 T_{R,L}=\f{T_- \mp T_+}{\O_-^1-\O_+^1}.
    \eea
\end{itemize}
We assume that the black hole entropies could be reproduced by the Cardy formula
\be \label{cardy} S_{R,L}=\f{\pi^2}{3}c_{R,L}^1 T_{R,L}^1, \ee
then the central charges of the right- and left-moving sectors could be derived as
\be \label{e7}
c_{R,L}^1=\f{3}{2\pi^2}\f{(\O_-^1-\O_+^1)(S_+ \mp S_-)}{T_- \mp T_+}.
\ee
It could be shown that if $T_+S_+=T_-S_-$ is satisfied then the right- and left-moving sector central charges are always equal $c_R^1=c_L^1$. We substitute $S_\pm=S_L \pm S_R$ into (\ref{e3}) and take variations on both sides of the equation while keeping $N_2$ invariant, then we get
\be \label{e63}
\f{\p \ma F}{\p N_1}d N_1=\f{S_L d S_L-S_R d S_R}{2\pi^2}.
\ee
From the above equation (\ref{e63}), using the Cardy formula (\ref{cardy}), the first law (\ref{e4}) and the fact that $c_R^1=c_L^1$, we obtain\footnote{Note that the extremal version of the formula has already appeared in \cite{Azeyanagi:2008dk}.}
\be \label{c1}
c_{R,L}^1=6\f{\p \ma F}{\p N_1},
\ee
which are the central charges for $N_1$ picture CFT. Therefore we see that if the entropy product function is mass-independent, the central charges of the dual CFT could be read easily from (\ref{c1}).

 Similarly, one could keep $N_1$ invariant and get the $N_2$ picture CFT with the temperatures and central charges
\bea \label{t2}
&& T_{R,L}^2=\f{T_- \mp T_+}{\O_-^2-\O_+^2},  \nn\\
&& c_{R,L}^2=\f{3}{2\pi^2}\f{(\O_-^2-\O_+^2)(S_+ \mp S_-)}{T_- \mp T_+},
\eea
and when $T_+S_+=T_-S_-$ the central charges could also be written as
\be \label{c2}
c_{R,L}^2=6\f{\p \ma F}{\p N_2}.
\ee
The equations (\ref{c1}) and (\ref{c2}) are convenient formulas of getting the central charges in the thermodynamics method when $S_+ S_-$ is mass-independent. Actually they could be generalized to other cases easily. We summarize the following important results:
\begin{itemize}
   \item The condition $T_+ S_+=T_-S_-$ is equivalent to the claim that $\ma F=\f{S_+ S_-}{4\pi^2}$ is mass-independent, and this also indicates that the right- and left-moving sector central charges  $c_{R,L}^{1,2}$ of the dual CFT are equal $c_R^{1}=c_L^{1}$, $c_R^{2}=c_L^{2}$.
   \item If $\ma F$ is mass-independent, it is the function of quantized charges $\ma F=\ma F(N_1,N_2, \cdots)$. Then with respect to every conserved charge $N_i$ , there could be a dual CFT picture with the central charges
       \be
       c^i_{R,L}=6\f{\p \ma F}{\p N_i}.
       \ee
       Note that the existence of the conserved charge does not guarantee that there is a corresponding CFT dual. For example, in the case of doubly rotating black ring, there are two $U(1)$ rotating symmetries, say $J_{\phi}$ and $J_\psi$, but there is no corresponding dual $J_\psi$ CFT, and this could be traced to the fact that $\ma F=J_\phi^2$ is independent of $J_\psi$ for this case \cite{Chen:2012yd}.
\end{itemize}

When $T_+S_+=T_-S_-$ is not satisfied, then we have $S_+S_-$ being mass-dependent.  We can still naively use (\ref{e7}) and get the central charges $c_{R,L}^1$, but now the right- and left-moving central charges do not equal $c_{R}^1 \neq c_{L}^1$. For the Einstein gravity whose action is classically diffeomorphism invariant, it is natural for such a black hole to have CFT duals with equal right- and left-moving central charges. So it was proposed in \cite{Chen:2012mh} that there are no CFT duals for the black holes with $T_+S_+\neq T_-S_-$. However, to our knowledge there is no proof that diffeomorphism of general relativity is quantum mechanically anomoly-free. Nevertheless, in the following discussion, we assume that the diffeomorphism is intact in the Einstein gravity so that the CFT candidates for a black hole should have equal left- and right-moving central charges.

Besides the above two elementary CFT pictures with the central charge $c_{R,L}^{1,2}$, there could be other CFT pictures generated by $SL(2,Z)$ group. We have got the $N_1$ picture CFT by considering the perturbations $dN_2=0$ in the thermodynamics (\ref{e8}) and got the $N_2$ picture CFT by considering the perturbations $dN_1=0$. In getting $N_1$ picture the perturbations are the ones $(d N_1,d N_2)=dN_1(1,0)$ with $dN_1$ being any small integers, and in getting $N_2$ picture the perturbations are the ones $(d N_1,d N_2)=dN_2(0,1)$ with $dN_2$ being any small integers. We may classify all the perturbations by two coprime integers $(a,b)$ by writing
$(d N_1,d N_2)=dN(a,b)$ with $dN$ being any small integers, and note that $(a,b)$ and $(-a,-b)$ denote the same kind of perturbations. We restrict the perturbations to be $(a,b)$ type $(d N_1,d N_2)=dN(a,b)$, then we could get the first laws (\ref{e8})
\bea
&&\f{1}{2} d M=T_R d S_R + \O_R^N d N \nn\\
&&\phantom{\f{1}{2} d M}=T_L d S_L + \O_L^N d N,
\eea
with the intensive quantities conjugate to the the integer $N$ being $\O_{R,L}^N=a \O_{R,L}^1 + b \O_{R,L}^2$. Using the above procedure, it can be shown easily one could get the $N$ picture CFT with the temperatures and the central charges
\bea \label{t3}
&& T_{R,L}^{(a,b)}=\f{1}{a/ T_{R,L}^1+b/T_{R,L}^2},  \nn\\
&& c_{R,L}^{(a,b)}=a c_{R,L}^1+b c_{R,L}^2.
\eea
Since the CFT pictures are in one to one correspondence with a pair of coprime integers $(a,b)$, we could call the $N_1$ picture CFT as $(1,0)$ picture, the $N_2$ picture as $(0,1)$ picture and general picture as $(a,b)$ picture. According to B\'ezout's lemma, for every pair of coprime integers $(a,b)$ there exists other pairs of coprime integers $(c,d)$ so that $ad-bc=1$, and note that the pair of coprime integers $(c,d)$ is not unique. So the $(a,b)$ picture CFT could be viewed as the $SL(2,Z)$ transformation of the elementary (1,0) and (0,1) pictures,
\be \label{e10}
\lt( \ba{c} c_{R,L}^{(a,b)} \\ c_{R,L}^{(c,d)} \ea \rt) = \lt(\ba{cc} a & b \\ c & d \ea \rt)
\lt( \ba{c} c_{R,L}^{(1,0)} \\ c_{R,L}^{(0,1)} \ea \rt), ~~~
\lt(\ba{cc} a & b \\ c & d \ea \rt) \in SL(2,Z).
\ee

We could actually get more from the thermodynamics \cite{Chen:2012ps}. In the (1,0) picture we have the perturbation $dM=\o, dN_1=k_1,dN_2=0$ around the black hole in the gravity side, and then we have the first laws (\ref{e8}) written as
\bea
&& T_R^1 d S_R= R_1 \lt( \f{1}{2}\o-\O_R k \rt),  \nn\\
&& T_L^1 d S_L= R_1 \lt( \f{1}{2}\o-\O_L k \rt).
\eea
In the CFT side, we suppose that
\bea
&& T_R^1 d S_R=\o_R^1-q_R^1 \m_R^1,  \nn\\
&& T_L^1 d S_L=\o_L^1-q_L^1 \m_L^1,
\eea
with $\o_{R,L}^1$, $q_{R,L}^1$ and $\m_{R,L}^1$ as the frequencies, charges, and chemical potentials of the perturbation around the thermal equilibrium. Explicitly we get the results
\bea \label{e31}
&& \o_{R,L}^1=\f{R_1}{2}\o=\f{T_-^2-T_+^2}{2T_- T_+(\O_-^1 - \O_+^1)}\o,  \nn\\
&& q_{R,L}^1=k_1,  \nn\\
&& \m_{R}^1=R_1 \O_{R}^1=\f{(T_- - T_+)(T_-\O_+^1 + T_+\O_-^1)}{2T_-T_+(\O_-^1 - \O_+^1)},  \nn\\
&& \m_{L}^1=R_1 \O_{L}^1=\f{(T_- + T_+)(T_-\O_+^1 - T_+\O_-^1)}{2T_-T_+(\O_-^1 - \O_+^1)}.
\eea
There are similar quantities in the $N_2$ picture CFT. For $(a,b)$ picture we have the perturbation $dM=\o, (dN_1,dN_2)=k(a,b)$ in the gravity side, and in the CFT side there is the perturbation with frequencies, charges, and chemical potentials
\bea \label{e33}
&& \o_{R,L}^{(a,b)}=\f{T_-^2-T_+^2}{2T_- T_+[(a\O_-^1+b\O_-^2) - (a\O_+^1+b\O_+^2)]}\o,  \nn\\
&& q_{R,L}^{(a,b)}=k,  \nn\\
&& \m_{R}^{(a,b)}=\f{(T_- - T_+)(T_-(a\O_+^1+b\O_+^2) + T_+(a\O_-^1+b\O_-^2))}{2T_-T_+[(a\O_-^1+b\O_-^2) - (a\O_+^1+b\O_+^2)]},  \nn\\
&& \m_{L}^{(a,b)}=\f{(T_- + T_+)(T_-(a\O_+^1+b\O_+^2) - T_+(a\O_-^1+b\O_-^2))}{2T_-T_+[(a\O_-^1+b\O_-^2) - (a\O_+^1+b\O_+^2)]}.
\eea

The description of the $SL(2,Z)$ generated pictures is a little different from the original one considered in \cite{Chen:2011wm,Chen:2011kt,Chen:2012mh}. In \cite{Chen:2012mh} the way of getting the general CFT duals is as what follows. One firstly redefines the quantum numbers by an $SL(2,Z)$
\be
\lt( \ba{c} N_{1'} \\ N_{2'} \ea \rt)=
\lt(\ba{cc} d & -c \\ -b & a \ea \rt)
\lt( \ba{c} N_{1} \\ N_{2} \ea \rt), ~~~
\lt(\ba{cc} d & -c \\ -b & a \ea \rt) \in SL(2,Z).
\ee
Then $N_{1',2'}$ will be good quantum numbers with the intensive quantities
\be
\lt( \ba{c} \O_\pm^{1'} \\ \O_\pm^{2'} \ea \rt)=
\lt(\ba{cc} a & b \\ c & d \ea \rt)
\lt( \ba{c} \O_\pm^{1} \\ \O_\pm^{2} \ea \rt).
\ee
Then the first laws (\ref{e9}) become
\bea
&&d M=T_+ d S_+ + \O_+^{1'} d N_{1'} + \O_+^{2'} d N_{2'}  \nn\\
&&\phantom{d M}=-T_- d S_- + \O_-^{1'} d N_{1'} + \O_-^{2'} d N_{2'}.
\eea
From the first laws one can set $dN_{2'}=0$, or $dN_{1'}=0$, and get the $N_{1'}$ or $N_{2'}$ picture central charges $c_{R,L}^{1'}$ or $c_{R,L}^{2'}$ with
\be \label{e11}
\lt( \ba{c} c_{R,L}^{1'} \\ c_{R,L}^{2'} \ea \rt) = \lt(\ba{cc} a & b \\ c & d \ea \rt)
\lt( \ba{c} c_{R,L}^{1} \\ c_{R,L}^{2} \ea \rt).
\ee
Note that these results are the same as  (\ref{e10}), and this is not simply a coincidence. In getting $c_{R,L}^{1'}$ (\ref{e11}) we are effectively considering the perturbations $dN_{1'}=d dN_1-c dN_2$, and in getting $c_{R,L}^{(a,b)}$ in (\ref{e10}) we are effectively considering the perturbations of the type $(dN_1,dN_2)=dN(a,b)$. Using $ad-bc=1$, we find that $dN_{1'}=dN$, which means that we are considering the variations of the same quantum number in two methods. Then it is of no wonder that they give the same central charges. Thus we have presented the equivalence of  two methods of getting general pictures.

\subsection{ASG analysis}

The shortage of ASG analysis of getting the black hole/CFT correspondence is that it applies only to the extremal black holes, and the advantage is that it is the only way to give a direct derivations of the central charges. Note that in the thermodynamics method or the hidden conformal symmetry we only have derivation of the temperatures, and the central charges have to be produced by the Cardy formula.

To get both the angular and charge pictures for the four-dimensional rotating charged black hole, we have to uplift the solution to five dimensions. Then we have 5D black hole solution
\be \label{e15}
ds^2_5=ds^2_4+\f{4\ell_p^2}{e^2} \lt( d\chi+e A \rt)^2
\ee
with $ds^2_4$ and $A$ denoting the 4D metric (\ref{ds4}) and gauge potential (\ref{e60}), $\chi \sim \chi+2\pi$, and $e$ being the unit charge. The geometry is the solution of pure Einstein gravity
\be \label{e14}
I_5=\f{1}{16\pi G_5} \int d^5x \sr{-G}(R_5-2\L),
\ee
with $G_5=\f{4\pi\ell_p}{e}G_4$. We define
\be \label{e23}
N^\chi=e A_\m\xi^\m=e(A_t-A_\phi N^\phi),
\ee
thus we could write
\be
d\chi+eA=d\chi+N^\chi dt+e A_\phi (d\phi+N^\phi dt).
\ee
Thus we could rewrite 5D metric (\ref{e14})  as
\bea \label{ds5}
&& ds^2_5=-N^2 dt^2+h_{rr}dr^2+h_{\th\th}d\th^2+ ( h_{\phi\phi}+4\ell_p^2 A_\phi^2 ) ( d\phi+N^\phi dt )^2 \nn\\
&& \phantom{ds^2_5=}  +\f{8\ell_p^2 A_\phi}{e} ( d\phi+N^\phi dt ) \lt( d\chi+N^\chi dt \rt)
                      +\f{4\ell_p^2}{e^2} ( d\chi+N^\chi dt )^2.
\eea
Similar embedding has been considered in \cite{Chen:2011wm} for the Kerr-Newman black hole where the explicit form of the metric was used. As stressed in \cite{Chen:2012ps}, the scale factor in (\ref{e15}) is chosen carefully, so that the angular momentum corresponding to the extra dimension $J_\chi=Q/e$ must be an integer. The angular velocities conjugate to $J_\chi$ at the outer and inner horizons are
\be
\O_\pm^\chi=-N^\chi|_{r=r\pm}=e\Phi_\pm.
\ee
We also denote $\O_\pm^\phi=\O_\pm$, $J_\phi=J$. One could show that the areas of the outer and inner horizons are
\be
(A_\pm)_5=\f{4\pi \ell_p}{e}A_\pm,
\ee
with $A_\pm$ being the horizon areas in four dimensions. Note  that the embedding does not affect the entropies of the horizons
\be
(S_\pm)_5=\f{(A_\pm)_5}{4G_5}=\f{A_\pm}{4G_4}=S_\pm.
\ee
Then there are first laws for the 5D rotating black hole
\bea
&&d M=T_+ d S_+ + \O_+^\phi d J_\phi + \O_+^\chi d J_\chi  \nn\\
&&\phantom{d M}=-T_- d S_- + \O_-^\phi d J_\phi + \O_-^\chi d J_\chi,
\eea
which are just (\ref{e9}) with $\O_\pm^{\phi,\chi}=\O_{\pm}^{1,2}$, $J_{\phi,\chi}=N_{1,2}$. Thus four-dimensional and five-dimensional black holes have the same first laws, and so they have the same CFT duals.

We take extremal limit of the metric (\ref{ds5}). Following \cite{Chen:2011wm}, we define
\be
f_3^{\phi,\chi}=\p_r N^{\phi,\chi},
\ee
and then in the extremal limit we have
\be
f_3^{\phi,\chi}|_{r=r_+}=\lim_{r_- \to r_+}\f{\O_-^{\phi,\chi}-\O_+^{\phi,\chi}}{r_+-r_-}.
\ee
We also define
\be
f_4=\lt.\f{f_2}{f_1}\rt|_{r=r_+},\hs{3ex} f^{\phi,\chi}=\lt.f_3^{\phi,\chi}f_4\rt|_{r=r_+}.
 \ee
 As stated before, according to the zeroth laws of the black hole, we must have $f_3^{\phi,\chi}|_{r=r_+}$, $f_4$ being constants, and so we have $f^{\phi,\chi}$ being constants. Thus we justify the assumptions in \cite{Chen:2011wm}. Since the 5D black hole is a solution of the Einstein gravity, thus we always have CFT duals for the extremal black hole with the central charges and the temperatures in the $J_{\phi}$ and $J_\chi$ CFT pictures
\be \label{e17}
c_L^{\phi,\chi}=\f{6 f^{\phi,\chi} S_+}{\pi}, ~~~ T_L^{\phi,\chi}=\f{1}{2\pi f^{\phi,\chi}}.
\ee
Upon a redefinition of the angles using an $SL(2,Z)$ transformation
\be
\lt( \ba{c} \phi' \\ \chi' \ea \rt) = \lt(\ba{cc} a & b \\ c & d \ea \rt)
\lt( \ba{c} \phi \\ \chi \ea \rt), ~~~
\lt(\ba{cc} a & b \\ c & d \ea \rt) \in SL(2,Z),
\ee
one gets $J_{\phi'}$ and $J_{\chi'}$ CFT pictures with the central charges and the temperatures
\bea
&& c_L^{\phi'}=a c_L^{\phi}+b c_L^{\chi}, ~~~ T_L^{\phi'}=\f{1}{a/T_L^{\phi}+b/T_L^{\chi}}  \nn\\
&& c_L^{\chi'}=c c_L^{\phi}+d c_L^{\chi}, ~~~ T_L^{\chi'}=\f{1}{c/T_L^{\phi}+d/T_L^{\chi}}.
\eea

On the other hand, taking the extremal limit of the right- and left-moving temperatures $T_{R,L}^{1}$ in (\ref{e16}) and the central charges $c_{R,L}^1$ in (\ref{t2}), and using the Hawking temperatures (\ref{ht}) and the definition of $f^\phi$, we find
\be
T_R^1=0, ~~~ T_L^1=\f{1}{2\pi f^\phi}, ~~~ c_{L}^1=\f{6 f^{\phi} S_+}{\pi},
\ee
which are in accord with (\ref{e17}). Similarly, the temperatures of the other pictures got from thermodynamics method could be shown to match the temperatures got here. Thus we have proved that for the 4D rotating charged black holes, the extremal limit of the CFT duals from the thermodynamics method always match the results of the ASG analysis of the extremal black hole. Since we did not use the explicit form of the solution, our derivations are quite general and could be applied to the other cases with multiple $U(1)$ charges.

\subsection{Hidden conformal symmetry}

To investigate the scattering of a complex scalar off the black hole, we can consider either a charged scalar in the 4D charged black hole background (\ref{ds4}), or equivalently a neutral scalar in the uplifted 5D black hole background (\ref{ds5}). In the former case, we consider a scalar of mass $\m_4$ and charge $k_e e$, with $e$ being the unit charge and $k_e$ being an integer. The equation of motion for such scalar $\Phi_4=e^{-i\o t+ik_\phi\phi}R(r)\Th(\th)$\footnote{Note that only when the scalar equation is separable could we write the scalar field in this way, but we write it in this way because the separability is always required for the existence of hidden conformal symmetry.} is
\be \label{e20}
(\na_\m-i k_e e A_\m)(\na^\m-i k_e e A^\m)\Phi_4=\m_4^2\Phi_4.
\ee
In the latter case we consider a scalar of mass $\m_{5}$, with its equation of motion being
\be \label{e21}
\na_M \na^M \Phi_5=\m_{5}^2\Phi_5.
\ee
We expand $\Phi_5=e^{ik_\chi\chi}\Phi_4=e^{-i\o t+ik_\phi \phi+ik_\chi\chi}R(r)\Th(\th)$, and then we can show that the equations (\ref{e20}) and (\ref{e21}) are identical if we have
\be
k_e=k_\chi, ~~~ \m_4^2=\m_{5}^2+\f{e^2 k_\chi^2}{16\pi \ell_{p}^2}.
\ee
The above identified equation is of the form
\bea \label{e68}
&& \f{1}{f_1 f_2 f_5 f_6} \p_r \f{f_1 f_5 f_6}{f_2}(r-r_+)(r-r_-)\p_r \Phi_4
+\f{[\o+N^\phi k_\phi+e(A_t-A_\phi N^\phi) k_e]^2}{f_1^2(r-r_+)(r-r_-)}\Phi_4 \nn\\
&& +\f{1}{f_1 f_2 f_5 f_6} \p_\th \f{f_1 f_2 f_6}{f_5}\p_\th \Phi_4-\f{(k_\phi-e A_\phi k_e)^2}{f_6^2}\Phi_4=\m^2_4 \Phi_4.
\eea

To use the hidden conformal symmetry, the first requirement is that the above equation can be separated. For a general black hole, the separability of the scalar equation is not guaranteed. The separability requires that
\be \label{e59}
\f{f_1 f_5 f_6}{f_2}=F(r) G(\th),
\ee
with $F(r)$ being some function of $r$ and $G(\th)$ being some function of $\th$. There is a similar requirement for $\f{f_1 f_2 f_6}{f_5}$, which will not be used here.
We can do integration on the both sides of the above equation (\ref{e59}) at the outer and inner horizons,
\be
\int d \th d \phi\lt. \frac{f_1 f_5f_6}{f_2} \rt|_{r=r_\pm} =F(r_{\pm})\int d \th d \phi G(\th).
\ee
Then using (\ref{ht}) we get
\be \label{e65}
F(r_{\pm})=\frac{\lambda}{r_+ -r_-}T_\pm A_\pm ,
\ee
with the constant $\lambda=\frac{4\pi}{\int d \th d \phi G(\th)}$ and $A_\pm$ being the areas of the outer and inner horizons. After variable separation we have the radial equation
\bea \label{r1}
&& \p_r F(r)(r-r_+)(r-r_-)\p_r R(r)
+\lt.\f{F(r)f_2^2}{f_1^2}\rt|_{r=r_+} \f{(\o-\O^\phi_+ k_\phi -\O^\chi_+ k_\chi)^2}{(r_+-r_-)(r-r_+)} R(r) \nn\\
&& -\lt.\f{F(r)f_2^2}{f_1^2}\rt|_{r=r_-} \f{(\o-\O^\phi_- k_\phi -\O^\chi_- k_\chi)^2}{(r_+-r_-)(r-r_-)} R(r)
=K_1(r)R(r).
\eea
Note that $k_e=k_\chi$, $K_1(r)$ depends on the explicit form of the metric. We only know that $K_1(r)$ is regular.

The hidden conformal symmetry requires moreover that we could redefine the radial coordinates $\r=\r(r)$ which is a monotonically increasing function of $r$ in the region $r \geq r_-$ such that
\be \label{e64}
F(r)(r-r_+)(r-r_-)\r'(r)=(\r-\r_+)(\r-\r_-),
\ee
with $\r_\pm \equiv \r(r_\pm)$.  $F(r)$ is neither vanishing nor divergent at $r \geq r_-$, but this does not guarantees that such a function $\r(r)$ exists. If such a radial coordinate $\r$ exists, by taking derivatives of $r$ on both sides of (\ref{e64}) and considering the values at the outer and inner horizons $r_\pm$, we have
\be \label{e26}
F(r_\pm)=\f{\r_+-\r_-}{r_+-r_-}.
\ee
From (\ref{e65})  this requires that the relation
\be
T_+A_+=T_-A_-
\ee
must be satisfied. For a black hole in Einstein gravity, the condition is just $T_+S_+=T_-S_-$ which has nontrivial implications in the thermodynamics method discussed before. It has been claimed to be the criterion for a black hole in the Einstein gravity to have CFT duals. We see that $T_+A_+=T_-A_-$ is the necessary condition that the hidden conformal symmetry could be found. Note that the condition is not sufficient, i.e.\! that from $T_+A_+=T_-A_-$ we cannot ensure the hidden conformal symmetry exists. In fact in the region of $r\geq r_+$, we always have
\be
\r(r)=\f{1}{2}\lt( \r_+ +\r_-+\f{\r_+-\r_-}{\tanh\lt[\f{\r_+-\r_-}{2} \int_r^\inf \f{dr}{F(r)(r-r_+)(r-r_-)}\rt]} \rt),
\ee
which is monotonically increasing. But in the region $r_- \leq r \leq r_+$, the formula is ill-defined, and analytical continuation is needed, but such continuation does not always exist.

Suppose that $T_+A_+=T_-A_-$ is satisfied and we have the required $\r(r)$. Then using (\ref{ht}), (\ref{e26}) we could write the radial equation (\ref{r1}) as
\bea \label{r2}
&& \p_\r(\r-\r_+)(\r-\r_-)\p_\r R(\r)
+\f{(\r_+-\r_-)(\o-\O_+^\phi k_\phi-\O_+^\chi k_\chi)^2}{16\pi^2T_+^2(\r-\r_+)} R(\r) \nn\\
&& -\f{(\r_+-\r_-)(\o-\O_-^\phi k_\phi-\O_-^\chi k_\chi)^2}{16\pi^2T_-^2(\r-\r_-)} R(\r)=K(\r) R(\r).
\eea
And again, $K(\r)$ is a regular function of $r$, and so a regular function of $\r$. We need $K(\r)$ to be a constant under appropriate approximations, but again this requirement is not always satisfied. However let us proceed with the assumption that $K$ can be set to be a constant.

In the study of the hidden conformal symmetry, the scalar radial equation  could be written in terms of the Casimir operator of the $SL(2,R)\times SL(2,R)$ symmetry $\ma H^2 R(\r)=K R(\r)$, with
\bea \label{r3}
&& \ma H^2=\p_\r(\r-\r_+)(\r-\r_-)\p_\r
+\f{(\r_+-\r_-)[\pi(T_L^C+T_R^C)\o+(n_L^C+n_R^C)k]^2}{16\pi^2 (T_L^C n_R^C-T_R^C n_L^C)^2 (\r-\r_+)}  \nn\\
&&\phantom{\ma H^2=}
-\f{(\r_+-\r_-)[\pi(T_L^C-T_R^C)\o+(n_L^C-n_R^C)k]^2}{16\pi^2 (T_L^C n_R^C-T_R^C n_L^C)^2 (\r-\r_-)}.
\eea
Here $K$ is a constant, $k$ is the quantum number of an angle that has not be assigned up to now and the letter $C$ in the superscript denotes the corresponding CFT. According to \cite{Chen:2010ni,Chen:2011kt}, the above radial equation can always be solved and gives the retarded Green's function and the absorption cross section that agree with the CFT predictions,
\bea
&&G_R \pp \sin \lt(\pi h_L^C+ i \f{ \o_L^C-q_L^C \m_L^C}{2T_L^C} \rt)
                          \sin \lt(\pi h_R^C+ i \f{ \o_R^C-q_R^C\m_R^C}{2T_R^C} \rt)  \nn\\
&&\phantom{G_R\pp} \times \G \lt( h_L^C-i \f{\o_L^C-q_L^C\m_L^C}{2\pi T_L^C} \rt)
                            \G \lt( h_L^C+i \f{\o_L^C-q_L^C\m_L^C}{2\pi T_L^C} \rt)  \nn\\
&&\phantom{G_R\pp} \times\G \lt( h_R^C-i \f{\o_R^C-q_R^C\m_R^C}{2\pi T_R^C} \rt)
                            \G \lt( h_R^C+i \f{\o_R^C-q_R^C\m_R^C}{2\pi T_R^C} \rt),
\eea
\bea
&&\s \pp \sinh \lt(\f{ \o_L^C-q_L^C \m_L^C}{2T_L^C}+\f{ \o_R^C-q_R^C\m_R^C}{2T_R^C} \rt)  \nn\\
&&\phantom{\s\pp} \times  \lt| \G \lt( h_L^C+i \f{\o_L^C-q_L^C\m_L^C}{2\pi T_L^C} \rt) \rt|^2
                           \lt| \G \lt( h_R^C+i \f{\o_R^C-q_R^C\m_R^C}{2\pi T_R^C} \rt) \rt|^2.
\eea
The conformal weights, frequencies, charges, and chemical potentials of the perturbation in the CFT could be identified as
\bea \label{e30}
&& h_{R,L}^C=\f{1}{2}\pm\sr{\f{1}{4}+K},  \nn\\
&& \o_{R,L}^C=\f{\pi T_L^C T_R^C \o}{T_L^C n_R^C-T_R^C n_L^C}, \nn\\
&& q_{R,L}^C=k, \\
&& \m_R^C=-\f{T_R^C n_L^C}{T_L^C n_R^C-T_R^C n_L^C}, \nn\\
&& \m_L^C=-\f{T_L^C n_R^C}{T_L^C n_R^C-T_R^C n_L^C}.   \nn
\eea

To get $J_\phi$ picture, or (1,0) picture, we consider the scattering of the scalar modes with $k_\chi=0$ in (\ref{r2}). Identifying the above two radial equations (\ref{r2}),(\ref{r3}), we find
\be \label{e29}
k=k_\phi, ~~~
T_{R,L}^\phi=\f{T_- \mp T_+}{\O_-^\phi - \O_+^\phi}, ~~~
n_{R,L}^\phi=-\f{\pi(T_-\O_+^\phi \mp T_+\O_-^\phi)}{\O_-^\phi - \O_+^\phi}.
\ee
We see the temperatures here got from the hidden conformal symmetry are the same as the ones got in the thermodynamics method (\ref{e16}). Moreover, using the results (\ref{e29}) and the formulas (\ref{e30}), we finally get the frequencies, the charges, and the chemical potentials of the perturbation in the CFT side, all of which are the same as the ones from the thermodynamics method (\ref{e31}). %This implies that from the hidden conformal symmetry we can read nothing but the thermodynamics of the black hole, including both the outer and inner horizons.

To get $J_\chi$, or (0,1), picture, we consider the scattering of the scalar modes with $k_\phi=0$ in (\ref{r2}). Identifying (\ref{r2}),(\ref{r3}), we find
\be
k=k_\chi, ~~~
T_{R,L}^\chi=\f{T_- \mp T_+}{\O_-^\chi - \O_+^\chi}, ~~~
n_{R,L}^\chi=-\f{\pi(T_-\O_+^\chi \mp T_+\O_-^\chi)}{\O_-^\chi - \O_+^\chi}.
\ee
Using the formulas (\ref{e30}), we could get the frequencies, charges, and chemical potentials of the perturbation in the CFT side. All the results are the same as the ones got in the thermodynamics method.

Also we can get the general $(a,b)$ picture with $a,b$ coprime integers. Let us consider the scattering of the scalar modes with $(k_\phi,k_\chi)=k(a,b)$ in (\ref{r2}). Identifying (\ref{r2}),(\ref{r3}), we find
\be \label{e36}
T_{R,L}^{(a,b)}=\f{T_- \mp T_+}{(a\O_-^\phi+b\O_-^\chi) - (a\O_+^\phi+b\O_+^\chi)}, ~~~
n_{R,L}^{(a,b)}=-\f{\pi[T_-(a\O_+^\phi+b\O_+^\chi) \mp T_+(a\O_-^\phi+b\O_-^\chi)]}
                   {(a\O_-^\phi+b\O_-^\chi) - (a\O_+^\phi+b\O_+^\chi)}.
\ee
Using the formulas (\ref{e30}), we obtain the frequencies, the charges, and the chemical potentials of the perturbation in the CFT side. The results are the same as the ones in the thermodynamics method (\ref{t3}), (\ref{e33}).

Just like in the thermodynamics method, the description here is in accord with the $SL(2,Z)$ generated pictures proposed in \cite{Chen:2011wm,Chen:2011kt}. According to \cite{Chen:2011wm,Chen:2011kt}, one redefines the angles
\be
\lt( \ba{c} \phi' \\ \chi' \ea \rt)=
\lt(\ba{cc} a & b \\ c & d \ea \rt)
\lt( \ba{c} \phi \\ \chi \ea \rt), ~~~
\lt(\ba{cc} a & b \\ c & d \ea \rt) \in SL(2,Z).
\ee
The quantity $k_\phi \phi+k_\chi \chi$ being invariant requires that the quantum numbers $k_{\phi,\chi}$ transform as
\be
\lt( \ba{c} k_{\phi'} \\ k_{\chi'} \ea \rt)=
\lt(\ba{cc} d & -c \\ -b & a \ea \rt)
\lt( \ba{c} k_{\phi} \\ k_{\chi} \ea \rt).
\ee
Using the hidden conformal symmetry, we could set $k_{\chi'}=0$ or $k_{\phi'}=0$, and get $J_{\phi'}$ or $J_{\chi'}$ pictures with the temperatures
\be \label{e37}
\lt( \ba{c} 1/T_{R,L}^{\phi'} \\ 1/T_{R,L}^{\chi'} \ea \rt) = \lt(\ba{cc} a & b \\ c & d \ea \rt)
\lt( \ba{c} 1/T_{R,L}^{\phi} \\ 1/T_{R,L}^{\chi} \ea \rt).
\ee
 These results are in accord with (\ref{e36}) $T_{R,L}^{\phi'}=T_{R,L}^{(a,b)}$. In getting $T_{R,L}^{\phi'}$ (\ref{e37}) we are effectively considering the scattering of the scalar modes $k_{\phi'}=d k_\phi-c k_\chi$, and in getting $T_{R,L}^{(a,b)}$ in (\ref{e36}) we are considering the scattering of the scalar modes of the type $(k_\phi,k_\phi)=k(a,b)$. Using $ad-bc=1$, we find that $k_{\phi'}=k$, which means that we are considering the scattering of the same scalar modes. It is not a surprise that two treatments  give the same results.

\section{Rotating Charged Black Holes in Arbitrary Dimensions}\label{s3}

In this section, we generalize the results in the previous section to the rotating black holes with arbitrary number of electric charges in arbitrary dimensions. The process is of no radical difference from the four-dimensional case, so it will be brief here.

\subsection{General black hole}

The action of the $d$-dimensional $U(1)^{n_c}$ Einstein-Maxwell theory could be written as
\be
I=\f{1}{16\pi G_d} \int d^d x\sr{-g}(R-2\L)-\f{1}{4\O_{d-2}} \int d^d x\sr{-g}F^a_{\m\n}F^{a\m\n},
\ee
where we set $c=\hbar=1$, $G_d=\ell_p^{d-2}$ and $a=1,2,\cdots,n_c$ with $n_c$ being the number of the $U(1)$ gauge fields. Firstly we write the metric as
\be\label{f1}
ds^2_d=-N^2dt^2+h_{rr}dr^2+h_{pq}d\th^p d\th^q +h_{mn}(d\phi^m+N^mdt)(d\phi^n+N^ndt),
\ee
Here we have $p,q=1,2,\cdots,n_\th$ and $m,n=1,2,\cdots,n_\phi$, with $n_\th=[\frac{d}{2}]-1$ being the number of polar angles and $n_\phi=[\frac{d-1}{2}]$ being the number of rotating angles. For simplicity we will denote $\p_p \equiv \p_{\th^p}$ and $\p_m \equiv \p_{\phi^m}$. We still assume the black hole has $\partial_t$ and $\partial_{m}$ isometries. We assume that the electromagnetic potentials have the form
\be A^a=A^a_tdt+ A^a_m d\phi^m, \ee
with $\partial_t A^a_\mu=\partial_{m} A^a_\mu=0$, where the summation of the index $m$ is indicated.

We suppose that there are two physical horizons located at $r_\pm$. The equations (\ref{e62}) are still applicable, with $f_1, f_2$ being regular functions of $r, \th^p$. The corresponding Hawking temperatures are still (\ref{ht}). The angular velocities  and the electric potentials of the outer and inner horizons are respectively
\bea \label{j63}
&& \O_\pm^m=-N^m |_{r=r_\pm},  \nn\\
&& \Phi^a_\pm = -A^a_\mu \xi^\mu|_{r=r_\pm},
\eea
with $\xi=\partial_t-N^m\partial_{m}$. The Bekenstein-Hawking entropies of the outer and inner horizons are
\be \label{f5}
S_\pm=\frac{A_\pm}{4G_d}=\frac{1}{4\ell_p^{d-2}}\int  d^{n_\th}\th d^{n_\phi}\phi f_5 f_6|_{r=r_\pm},
\ee
with $f_5\equiv \sqrt{\det h_{pq}}$ and $f_6\equiv \sqrt{\det h_{mn}}$, which are regular positive functions of $r, \th^p$.

The first laws of thermodynamics at two horizons are,
\begin{eqnarray} \label{f2}
&& dM=T_+ dS_+ +\O^m_{+}dJ_m + \Phi^a_+ dQ_a \nn\\
&&\phantom{dM} =-T_- dS_- +\O^m_{-}dJ_m + \Phi^a_+ dQ_a,
\end{eqnarray}
with $J_m$ and $Q_a$ being the angular momenta and electric charges of the black hole.

\subsection{Thermodynamics method}

Firstly, we rewrite the first law (\ref{f2}) in terms of  the quantized parameters,
\bea \label{z58}
&&d M=T_+ d S_+ + \O_+^m d N_m + \O_+^a d N_a \nn\\
&&\phantom{d M}=-T_- d S_- + \O_-^m d N_m + \O_-^a d N_a,
\eea
with $N_m \equiv J_m$ and $N_a\equiv \frac{Q_a}{e_a}$ are integers. The parameter $e_a$ is the unit charge of a single particle corresponding to the gauge field $A^a$. Equivalently we could write the first laws as
\bea \label{e66}
&&d M=T_+ d S_+ + \O_+^i d N_i \nn\\
&&\phantom{d M}=-T_- d S_- + \O_-^i d N_i,
\eea
with the index $i=(m,a)$ and $i$ could take $n=n_\phi+n_c$ independent values. If $T_+S_+=T_-S_-$ is satisfied, the entropy product $\mathcal{F}= \frac{S_+S_-}{4\pi^2}$ is mass independent, then the thermodynamics method can be applied to discuss the CFT duals of the black hole if the black hole is the solution of the Einstein gravity or other gravity without diffeomorphism anomaly.

Using the thermodynamics method, one can set all the charges and momenta fixed except one $N_i$ and get the central charges referring to the $N_i$ picture
\be
c^{i}_{R,L}=6\frac{\partial\mathcal{F}}{\partial N_{i}}.
\ee
Also one could consider the perturbations of the type $dN_i=dN v_i$, with $\{v_i, i=1,2,\cdots,n\}$ being $n$ coprime integers. This gives the general picture characterized by the vector ${\bf v}=(v_1,v_2,\cdots,v_n)$ with central charges
\be \label{e69}
c_{R,L}^{\bf v}=v_i c_{R,L}^i,
\ee
and here a summation of the index $i$ is indicated. In the 4D case, the angular momentum and the electric charge form a vector space $Z^2$, and each vector $(a,b)$, with $a,b$ being coprime, of this space corresponds to a dual CFT picture. In the general case here, all the angular momenta and electric charges form a vector space $Z^n$, and each vector ${\bf v}=(v_1,v_2,\cdots,v_n)$, with $\{v_1,v_2,\cdots,v_n\}$ being coprime, corresponds to ${\bf v}$ picture CFT.

The prescription here is that the general ${\bf v}$ picture appears as the response of the black hole with respect to the perturbations characterized by ${\bf v}$.  In \cite{Chen:2012mh} it was proposed that the general picture appears as $SL(n,Z)$ redefinitions of the charges. Similar to the case with one angular momentum and one electric charge, the two prescriptions here are equivalent.
To show this we have to use the theorem that is proved in the appendix. We call a vector ${\bf v}=(v_1,v_2,\cdots,v_n)$ in the space $Z^n$ as the coprime vector if $\{v_1,v_2,\cdots,v_n\}$ being coprime integers. The theorem states that a coprime vector ${\bf v}=(v_1,v_2,\cdots,v_n)$ in space $Z^n$ is one-to-one correspondence to a row, or a column, of an $SL(n,Z)$ matrix. According to \cite{Chen:2012mh}, we define the charges $N_{i'}=N_i \L^i_{\phantom{i}i'}$ with $\L^i_{\phantom{i}i'}\in SL(n,Z)$. We define the inverse of $\L^i_{\phantom{i}i'}$ as $\L^{i'}_{\phantom{i'}i}$, then there is a CFT picture with the central charges
\be \label{z57}
c_{R,L}^{i'}=\L^{i'}_{\phantom{i'}i}c_{R,L}^{i}.
\ee
Since the coprime vector ${\bf v}$ is in one-to-one correspondence with, say, the $i'$-th row of the $SL(n,Z)$ matrix $\L^{i'}_{\phantom{i'}i}$, i.e. that $v_i=\L^{i'}_{\phantom{i'}i}$ with $i'$ being fixed and $i=1,2,\cdots,n$, and then (\ref{e69}) and (\ref{z57}) give the same central charges for ${\bf v}$ picture.

It is tempting to apply the thermodynamics method to general nonextremal black holes in string theory, for example the 5D and 4D rotating multi-charged black holes found in \cite{Cvetic:1996xz,Cvetic:1996kv} and more general black rings than the ones considered in \cite{Chen:2012yd}. This  will be reported in another place \cite{Zhang:2013}.

\subsection{ASG analysis}

For the general case with multiple charges here, we could uplift the $d$-dimensional black hole to $d+n_c$ dimensions. For each $U(1)$ charge, we assign one angular momentum corresponding to an extra dimension. The uplifted metric has the following form
\be \label{f3}
ds^2_{d+n_c}=ds^2_d+\sum^{n_c}_{a=1}\frac{16\pi \ell^{d-2}_{p}}{\Omega_{d-2}e^2_a}(d\chi^a+e_a A^a)^2,
\ee
with $ds^2_d$ denoting the $d$-dimensional metric (\ref{f1}), and $e_a$ being the unit charge corresponding to the gauge field $A^a$. The geometry is the solution of the pure $d+n_c$ dimensional Einstein gravity,
\be
I_{d+n_c}=\frac{1}{16\pi G_{d+n_c}}\int d^{d+n_c}x \sqrt{-G} (R_{d+n_c}-2\L).
\ee
with the Newton constant
$G_{d+n_c}=\frac{8^{n_c}\pi^{\frac{3n_c}{2}}\ell_p^{\frac{(d-2)n_c}{2}}}{e_1e_2\cdots e_{n_c} \Omega_{d-2}^{\frac{n_c}{2}}}G_d$.
We define
\be\label{Naa}
N^{a}=e_a(A^a_t-A^a_mN^m),
\ee
then we could write
\be
d\chi^a+e_a A^a=d\chi^a+ N^adt+e_aA^a_m(d\phi^m+N^mdt).
\ee
The metric (\ref{f3}) can be written as
\begin{eqnarray} \label{f6}
&& ds^2_{d+n_c}=-{N}^2dt^2+{h}_{rr}dt^2+{h}_{pq}d\th^p d\th^q+{\hat h}_{mn}(d\phi^m+N^mdt)(d\phi^n+N^ndt)  \nn\\
&& \phantom{ds^2_{d+n_c}=} +{\hat h}_{ma}(d\phi^m+N^mdt)(d\chi^a+N^adt)+{\hat h}_{ab}(d\chi^a+N^adt)(d\chi^b+N^bdt)
\end{eqnarray}
with the components of the metric being defined as
\begin{eqnarray}
&& {\hat h}_{mn} = h_{mn}+\frac{16\pi \ell^{d-2}_{p}}{\Omega_{d-2}}\sum_a A^a_m A^a_n , \nn\\
&& {\hat h}_{ma} = \frac{16\pi \ell^{d-2}_{p}}{\Omega_{d-2}e_a}A^a_m , \nn\\
&& {\hat h}_{ab} = \frac{16\pi \ell^{d-2}_{p}}{\Omega_{d-2}e^2_a} \delta_{ab}.
\end{eqnarray}
With this embedding, the charge referring to each $U(1)$
gauge field is presented by the corresponding angular momentum $J_a=\frac{Q_a}{e_a}$, which must be an integer. The angular
velocities conjugate to $J_a$ at the outer and inner horizons are
\be
\Omega_{\pm}^a=-N^a|_{r=r_\pm}=e_a\Phi^a_{\pm}
\ee
The areas of the outer and inner horizons for  two black holes have the relation
\be \label{f7}
(A_\pm)_{d+n_c}=\frac{8^{n_c}\pi^{\frac{3n_c}{2}}\ell_p^{\frac{(d-2)n_c}{2}}}{e_1e_2\cdots e_{n_c} \Omega_{d-2}^{\frac{n_c}{2}}}(A_\pm)_d,
\ee
with $(A_\pm)_d$ being the horizon areas in $d$ dimensions. Similarly the embedding does not change the entropies of the horizons,
\be
(S_\pm)_{d+n_c}=\frac{(A_\pm)_{d+n_c}}{4G_{d+n_c}}=\frac{(A_\pm)_d}{4G_d}=(S_\pm)_d .
\ee
The first laws for the $(d+n_c)$-dimensional rotating black hole are
\begin{eqnarray}
&& dM = T_+ dS_+ +\Omega^i_+dJ_i \nn\\
&& \phantom{dM} = -T_- dS_- +\Omega^i_- dJ_i,
\end{eqnarray}
which are just (\ref{e66}). Here we use the index $i=(m,a)$ which takes $n=n_\phi+n_c$ values.

To do ASG analysis we have to take extremal limit of the metric (\ref{f6}). We define
\be
f_3^{i}=\partial_r N^{i},
\ee
and in the extremal limit we have
\be
f_3^{i}|_{r=r_+}=\lim_{r_-\rightarrow r_+} \frac{\Omega_{-}^{i}-\Omega_{+}^{i}}{r_+-r_-}.
\ee
We also define $f_4\equiv\frac{f_2}{f_1}|_{r=r_+}$, $f^{i}\equiv f_3^{i}f_4|_{r=r_+}$, and the values $f_3^{i}|_{r=r_+}$, $f_4$, and $f^{i}$ are all constants. Then we have CFT duals of the extremal black hole with the central charges and the temperatures in the $J_i$ picture CFT
\be
c_L^{i}=\frac{6f^{i}S_+}{\pi},~~ T_L^{i}=\frac{1}{2\pi f^{i}}.
\ee
The temperature is in accord with the one from the thermodynamics method.

Upon redefinitions of the angles using an $SL(n,Z)$ transformation
\be
\phi^{i'}=\L^{i'}_{\phantom{i'}i}\phi^i, ~~~ \L^{i'}_{\phantom{i'}i} \in SL(n,Z),
\ee
one gets $J_{i'}$ picture with the central charges and the temperatures
\be
{c}_L^{i'}= \L^{i'}_{\phantom{i'}i} {c}_L^{i}, ~~~ {T}_L^{i'}=\f{1}{ \L^{i'}_{\phantom{i'}i}/T_L^{i}},  \nn\\
\ee
with the summation of the index $i$ in both equations. Therefore the thermodynamics method and the ASG method are consistent with each other for the extremal black hole of a general form.

\subsection{Hidden conformal symmetry}

To use the hidden conformal symmetry, we can either discuss the scattering of a charged scalar particle in the $d$-dimensional charged
black hole background, or equivalently a neutral scalar particle in the uplifted $(d+n_c)$-dimensional black hole background.
In the former case, the equation of motion is
\be \label{CG1}
(\nabla_\mu-i\sum_a k_ae_aA_\mu^a)(\nabla^\mu-i\sum_b k_b e_b A^{b\mu}) \Phi_d= \mu_d^2 \Phi_d,
\ee
where $\Phi_d=e^{-i\o t+ik_m \phi^m}R(r)\Theta(\th)$ is the wave function of the charged probe particle, $\Theta(\th)$ is a function of $\th^p$ with $p=1,2,\cdots,n_\th$ and $\mu_d$ is the mass of the probe. In the latter case, the equation of motion is$\phantom{\ref{诡异}}$
\be \label{j58}
\nabla_M\nabla^M \Phi_{d+n_c}= \mu_{d+n_c}^2 \Phi_{d+n_c},
\ee
where the wave function $\Phi_{d+n_c}$ describes a neutral scalar particle with mass $\mu_{d+n_c}$. We assume that the two wave functions are related by $\Phi_{d+n_c}=\Phi_{d} e^{i k_a\chi^a}$.  The equations (\ref{CG1}) and (\ref{j58}) are identical if we have
\be
\mu^2_{d}=\mu^2_{d+n_c}+\sum_{a}\frac{\Omega_{d-2}e_a^2}{16\pi \ell_p^{d-2}}k_a^2.
\ee
Then the equation of motion (\ref{CG1}) has the form
\bea \label{e67}
&&\frac{1}{f_1f_2f_5f_6}\partial_r \frac{f_1f_5f_6}{f_2}(r-r_+)(r-r_-)\partial_r \Phi_d
  +\frac{(\omega+k_i N^i)^2}{f_1^2(r-r_+)(r-r_-)} \Phi_d  \nn\\
&& + \frac{1}{f_1f_2f_5f_6} \partial_p f_1 f_2 f_5 f_6 h^{pq} \partial_q \Phi_d
 -h^{mn}(k_m-\sum_a k_ae_aA^a_m)(k_n-\sum_b k_b e_b A^b_n)\Phi_d=\mu^2_d\Phi_d.
\eea

The equation (\ref{e67}) is of no radical difference from its 4D cousin (\ref{e68}), so the calculations there could be generalized here straightforwardly. From the separability of (\ref{e67}) and the existence of well-defined conformal coordinate $\r(r)$ we could get the necessary condition for the existence of hidden conformal symmetry
\be
T_+A_+=T_-A_-,
\ee
with $T_\pm, A_\pm$ being the Hawking temperatures and the areas of the outer and inner horizons. Note that treating $A_\pm$ as horizon areas of the $d$-dimensional or $(d+n_c)$-dimensional black hole does not make any difference because of (\ref{f7}).

The equations (\ref{e59}) and (\ref{e64}) are still applicable here given that the definitions $f_{4,5}$ are different and $\th$ denotes $n_\th$ coordinates $\th^p$. When the black hole is asymptotically flat, in the limit $r\to\inf$ we have $f_1 \sim \f{1}{r}, f_2 \sim r, f_5 f_6 \sim r^{d-2}$, from (\ref{e59}) we get $F\sim r^{d-3}$, and then from (\ref{e64}) we get $\r\sim r^{d-3}$. Similarly for the black hole in AdS spacetime, in the limit $r\to\inf$ there are $f_{1,2} \sim 1, f_5 f_6 \sim r^{d-2}, F\sim r^{d-2}$, and then we get $\r\sim r^{d-1}$. This justifies the choices of conformal coordinates $\r=r^2$ for 3D BTZ black hole \cite{Banados:1992gq}, $\r=r$ for 4D Kerr black hole \cite{Castro:2010fd}, $\r=r^2$ for 5D Myers-Perry black hole \cite{Krishnan:2010pv}, and $\r=r^{d-3}$ for RN black hole in $d$ dimensions \cite{Chen:2012ps}.

Given that there is hidden conformal symmetry, we could get the $J_i$ (it may be $J_m$ or $Q_a$) picture CFT by considering the modes of the scalar with $k_j=0$ for all indexes $j\neq i$. For each $J_i$ picture, we have the temperatures
\be
T^{i}_{R,L}=\frac{T_-\mp T_+}{\Omega^{i}_--\Omega^{i}_+}.
\ee
Also, we can get the general ${\bf v}=(v_1,v_2,\cdots,v_n)$ picture by considering the modes of the scalar field with
$(k_1,k_2,\cdots,k_{n})=k{\bf v}$, and the temperatures are
\be
T^{{v}}_{R,L}=\frac{T_-\mp T_+}{v_i \Omega^{i}_-- v_i \Omega^{i}_+}.
\ee
Using the theorem proved in the appendix, we could show that the way of getting the general CFT pictures by considering different perturbations here is equivalent to the way of making $SL(n,Z)$ redefinitions of the quantum numbers $\{k_1,k_2,\cdots,k_n\}$. All these results are the same as the ones got from the thermodynamics method.

\section{Conclusion and Discussion}\label{s4}

In this paper we discussed the relation of the thermodynamics method with other conventional ways in setting up the black hole/CFT correspondence. We applied the thermodynamics method to find the CFT duals for the black holes in arbitrary dimensions and with arbitrary number of electric charges. We started from a metric of a general ADM form and assumed the existence of rotating symmetries and timelike Killing symmetry outside the black hole horizons. This would include large classes of black objects we have known, not only the rotating black holes but also black rings. We found that even without knowing the explicit forms of black holes,  the thermodynamics method could give consistent results with the ones obtained from ASG analysis and the hidden conformal symmetry.

In \cite{Chen:2012mh}, it was argued that $T_+S_+=T_-S_-$ is the necessary condition for a black hole in the Einstein(-Maxwell) gravity, or other gravity without diffeomorphism anomaly, to have CFT duals. The naively use of the thermodynamics method for a black hole for which $T_+S_+\neq T_-S_-$ would lead to unequal right- and left-moving central charges, which seems pathological in a theory that is diffeomorphism invariant.
In this paper, from the separability of the scalar equation and the existence of well-defined conformal coordinate it was shown that $T_+A_+=T_-A_-$ is the necessary condition for a black hole to have the hidden conformal symmetry. For the Einstein(-Maxwell) gravity, $T_+A_+=T_-A_-$ is just $T_+S_+=T_-S_-$.
Then according to the check in \cite{Chen:2012mh,Chen:2012ps}, for the Myers-Perry black holes in dimensions $d\geq6$, the Kerr-AdS black holes in dimensions $d\geq4$, and the RN-AdS black holes in dimensions $d\geq4$, $T_+A_+=T_-A_-$ is not satisfied so there is no hidden conformal symmetry for these black holes. The absence of the hidden conformal symmetry and the fact that the possible CFT has different left- and right-moving central charges both suggest that for these black holes there is no CFT dual picture. Nevertheless, the holographic pictures of extremal Kerr-AdS(-dS) black holes have been investigated via ASG analysis \cite{Lu:2008jk,Hartman:2008pb,Chen:2011wm} and superradiant scattering \cite{Chen:2010bh}. Even in these cases the extreme limit of the thermodynamics method still gives the consistent results. Note that in the extreme limit, the holographic CFT seems to be chiral with the right-moving temperature vanishing and right-moving central charge uncertain.

On the other hand, although there is $T_+S_+=T_-S_-$ for five-dimensional black rings, but the scalar equation cannot be separated, so there is  no hidden conformal symmetry in this case \cite{Chen:2012yd}. Nevertheless the holographic description could be established in this case, showing the power of the thermodynamics method.  It was shown in this paper, when the hidden conformal symmetry could be used, it always gives the same results with the ones got from the thermodynamics method.

In (\ref{ht}) and (\ref{j63}), we have used the formulas of calculating the angular velocities and electric potentials of the outer and inner horizons as
\bea \label{e61}
&&\O_\pm^m=-N^m |_{r=r_\pm},  \nn\\
&&\Phi_\pm^a=-A_\mu^a \xi^\mu|_{r=r_\pm}.
\eea
Actually they are only correct for the black holes in asymptotical flat spacetimes, and are not correct for Kerr-AdS and RN-AdS black holes in dimensions $d\geq4$. In the latter cases the formulas should be changed to \cite{Caldarelli:1999xj,Gibbons:2004ai}
\bea
&& \O_\pm^m= N^m |_{r\to \inf}  -  N^m |_{r=r_\pm},  \nn\\
&& \Phi_\pm^a= A_\mu^a \xi^\mu|_{r \to \inf}  -  A_\mu^a \xi^\mu|_{r=r_\pm}.
\eea
This modification dose not change the CFT temperatures (\ref{e16}), the central charges (\ref{e7}) and the frequencies (\ref{e31}), as the angular velocities always appear in the form $\O_-^m-\O_+^m$ in our discussion. But for the chemical potential of the possible CFT (\ref{e31}), there should be modification for Kerr-AdS and RN-AdS black holes in dimensions $d\geq4$. However due to the absence of the CFT dual in these cases, we need not worry about the implications of the modifications.

We have shown that  $T_+A_+=T_-A_-$ is the necessary condition for the hidden conformal symmetry, but it is not a sufficient condition. First of all the scalar equation should be separable, secondly there should be well-defined conformal coordinate $\r(r)$ in all regions $r \geq r_-$, and finally the function $K(\r)$ in (\ref{r2}) could be set to a constant under suitable approximations. Given we have all these conditions and the hidden conformal symmetry exists for a black hole, then the results from hidden conformal symmetry are the same with the ones got from the thermodynamics method and ASG analysis. The consistency of the three methods is in accord with the checks in \cite{Chen:2012mh,Chen:2012yd,Chen:2012ps} for concrete examples.

For a black hole in the Einstein(-Maxwell) gravity, we have used the condition $T_+S_+=T_-S_-$, or equivalently the mass-independence of the entropy product $S_+S_-$,  as the criterion whether the thermodynamics method could be used. But for a black hole in a gravity theory that is not diffeomorphism invariant, for example three-dimensional topologically massive gravity (TMG), one should not expect that $S_+S_-$ is mass-independent, and this is indeed the case as checked in \cite{Detournay:2012ug}. For these black holes the necessary condition for the existence of hidden conformal symmetry is just $T_+ A_+=T_-A_-$. In these cases the thermodynamics method is still applicable and gives the same results as the hidden conformal symmetry. The results in this case and some other investigations will be reported in another paper \cite{Chen:2013aza}.

In this paper, we discussed the black holes in Einstein(-Maxwell) gravity theories. It would be interesting to consider the gravity with scalar matter or the gravity with higher curvature corrections. In the latter case, there have been lots of study of the extremal black holes, in which cases the attractor mechanism has been developed to compute the entropy even though the explicit forms of the solutions are not clear. There are also multi-centered black holes in this case. How to apply the thermodynamics method
to these cases is an interesting issue.

One subtle issue in Kerr/CFT and its extensions is on the validity of Cardy formula. Actually this has been a long-existing question since the seminal work in \cite{Strominger:1996sh}. For higher dimensional black holes, whatever the black holes in string theory or just Kerr black hole, the central charges of their dual CFT depends on the quantized charges of the black holes. It is often the case that the central charges is large but the levels of excitation in the CFT is low. Therefore it is beyond the validity of the Cardy formula, which requires a well-defined saddle point approximation.  For the black holes in string theory, it has been argued that one could find a frame such that the Cardy formula could be applied. For the case in \cite{Strominger:1996sh}, such a frame with central charge $c=6$ have been shown to exist. For the multi-charged black holes, such a trick could be workable. However for a pure Kerr black hole, even if it were successfully embedded into string theory, we could not figure a way to solve this issue.

There is another peculiar thing on the CFT dual of the black holes in $D\geq 4$. For the BTZ black hole, it dual CFT has central charge $c=\frac{3\ell}{2G}$, which is only dependent of the AdS$_3$ radius or equivalently the cosmological constant. No matter how the black hole change, the central charge is intact. However, for the black holes in higher dimensions, the central charges of dual CFT are the functions of the charges of the black holes, angular momentum or U(1) charges. When the black hole evolves, say via Hawking radiation, the universal properties of dual CFT change accordingly. In particular, the change of the central charge indicates that there is a underlying RG flow. There is also the issue of matching the numbers of independent parameters for the black hole and the CFT. For the CFT there are three independent parameters, i.e. $c=c_R=c_L$, $T_R$ and $T_L$, however for the black hole the number of independent parameters is not necessarily three. For example, for Kerr or RN black hole the number is two, and for multicharged black holes or black rings the number could be larger than three. It is certainly valuable to have a better understanding of these issues\footnote{We would like to thank the anonymous referee to inspire this discussion.}.

\vspace*{10mm}
\noindent {\large{\bf Acknowledgments}}\\
The work was in part supported by NSFC Grant No. 10975005, 11275010. JJZ was also in part supported by Scholarship Award for Excellent Doctoral Student granted by Ministry of Education of China.
\vspace*{5mm}

\begin{appendix}

\section{Coprime Integers and $SL(n,Z)$ Matrix}

In the appendix, we give a simple proof of the theorem we have used in the paper. We have to say that we are not aware if it has appeared in the mathematical literature, but as we need it we just give a proof here. Note that the theorem is a generalization of the B\'ezout lemma, and so could be viewed as a proof of B\'ezout lemma.

For convenience, let us firstly make some simple definitions. Firstly we clarify the notion of coprime integers. We consider a vector ${\bf v}=(v_1,v_2,\cdots,v_n)$ in the space $Z^n$ with $n$ being a positive integer, which means that all $\{v_i, i=1,2,\cdots,n\}$ are integers.
\begin{dfn}[Coprime Integers]
We say that a finite number of integers $\{v_i\}$ are {\it coprime}, if their greatest common divisor being one.
\end{dfn}
\noindent We stress that the set of $n$ integers $\{v_i\}$ are coprime does not means that the integers of a subset are coprime. For example $\{ 2,4,5 \}$ are coprime but $\{ 2,4 \}$ are not.
\begin{dfn}[Coprime Vector]
 We say that the vector ${\bf v}=(v_1,v_2,\cdots,v_n)$ in space $Z^n$ is an {\it coprime vector}, if the $n$ integers $\{v_i\}$ are coprime.
\end{dfn}
\noindent Note that the $n$ zeroes $\{v_i=0,i=1,2,\cdots,n\}$ are not coprime, so the vanishing vector is not coprime. We know for a matrix there are three classes of elementary row operations:{\it row switching}, which means a row within the matrix can be switched with another row; {\it row multiplication}, which means each element in a row is multiplied by a non-zero constant; and {\it row addition}, which means a row is replaced by the sum of that row and a multiple of another row. There are three classes of elementary column operations similarly. Since we are only interested in the $SL(n,Z)$ matrices in this appendix, what we mean multiplication here is restricted to the multiplication of the integers. Also, here we only care about the elementary operations that do not change the determinant or the absolute value of determinant of a matrix, and so we define the so-called unitary elementary operation and anti-unitary elementary operation.
\begin{dfn}[Unitary Elementary Operation]
A unitary elementary operation is an elementary operation that does not change the determinant of a nondegenerate matrix.
\end{dfn}
\begin{dfn}[Anti-unitary Elementary Operation]
An anti-unitary elementary operation is an elementary operation that changes the determinant of a nondegenerate matrix by a single factor $-1$.
\end{dfn}
\noindent The unitary elementary operations include row addition and column addition. The anti-unitary elementary operations include row switching, row multiplication by $-1$, column switching and column multiplication by $-1$. Finally we denote the $n \times n$ anti-special linear transformation matrix in the domain $Z$ as $AL(n,Z)$ matrix.
\begin{dfn}[$AL(n,Z)$ Matrix]
An $AL(n,Z)$ matrix is a $GL(n,Z)$ matrix whose determinant is $-1$.
\end{dfn}
\noindent Note that an anti-unitary elementary operation changes an $SL(n,Z)$ matrix to an $AL(n,Z)$ matrix, and vise versa.

With the definitions above we can state the theorem as follows.
\begin{thm}\label{z56}
A coprime vector ${\bf v}=(v_1,v_2,\cdots,v_n)$ in space $Z^n$ is one-to-one correspondence to a row, or a column, of an $SL(n,Z)$ matrix $\L^{i}_{\phantom{i}j}$.
\end{thm}
\noindent Explicitly, the theorem states that for a coprime vector ${\bf v}=(v_1,v_2,\cdots,v_n)$ there exists $SL(n,Z)$ matrix $\L^{i}_{\phantom{i}j}$ which has ${\bf v}$ as, without loss of generality, its first row, i.e. that $\L^{1}_{\phantom{1}i}=v_i, i=1,2,\cdots,n$, and reversely every row, or column, of an $SL(n,Z)$ matrix constitute a coprime vector.

The proof of the second part of the theorem is simple. The determinant of the $SL(n,Z)$ matrix $\L^{i}_{\phantom{i}j}$ could be expanded by its every row (or column), say the first row, which means that there exist integers $k^i, i=1,2,\cdots,n$ satisfying
\be
\L^{1}_{\phantom{1}i}k^i=1.
\ee
This shows that the $n$ integers $\{\L^{1}_{\phantom{1}i}\}$ are coprime, and then they could constitute a coprime vector.

The proof of the first part of the theorem needs some labor. We begin by considering two vectors ${\bf v}=(v_1,v_2,\cdots,v_n)$ and $\td{\bf v}=(\td v_1,\td v_2,\cdots,\td v_n)$ that are related through some $n-1$ integers $k_i, i=2,3,\cdots,n$ as
\bea \label{a53}
&& \td v_1=v_1,  \nn\\
&& \td v_i=v_i+k_i v_1, ~~~ i=2,3,\cdots,n.
\eea
This can also be recast as
\bea \label{a54}
&& v_1=\td v_1,  \nn\\
&& v_i=\td v_i-k_i \td v_1, ~~~ i=2,3,\cdots,n.
\eea
The relation (\ref{a53}) shows that if ${\bf v}$ is not coprime, neither is $\td {\bf v}$, and (\ref{a54}) shows that if $\td {\bf v}$ is not coprime, neither is ${\bf v}$. Then we have the following lemma.
\begin{lmm} \label{z1}
For two vectors ${\bf v}$ and $\td {\bf v}$ in $Z^n$ related through (\ref{a53}) by some vector $\bf{k}$ in $Z^{n-1}$, that ${\bf v}$ is coprime is equivalent to that $\td {\bf v}$ is coprime.
\end{lmm}

Given a coprime vector ${\bf v}$ we want an $SL(n,Z)$ matrix with ${\bf v}$ being its first row, and when we have an $AL(n,Z)$ matrix with ${\bf v}$ being its first row, we can just make an anti-unitary elementary operation with the first row intact to change the $AL(n,Z)$ matrix to an $SL(n,Z)$ matrix. Equipped with the knowledge, we describe the procedure of \emph{substraction}.
\begin{enumerate}
  \item[I] If any element of the coprime vector ${\bf v}=(v_1,v_2,\cdots,v_n)$ is minus, change it to its opposite.
  \item[II] We choose the element of a coprime vector ${\bf v}$ in $Z^n$ that has the smallest but nonvanishing absolute value. If there are more than one choices, we just choose one of them. If the element is not $v_1$ we exchange the element with $v_1$.
  \item[III] The other values of ${\bf v}$, namely $v_i, i=2,3,\cdots,n$, are either vanishing, equal, or larger than $v_1$. Then for every $v_i, i=2,3,\cdots,n$, there exists unique $k_i$ that makes the value $\td v_i=v_i+k_i v_1$ satisfying $0\leq\td v_i \leq v_1-1$. We choose $\td v_1=v_1$. Because the vector ${\bf v}$ is coprime, according to Lemma~\ref{z1}, the vector $\td {\bf v}=(\td v_1,\td v_2,\cdots,\td v_n)$ is also coprime. Then as long as $v_1 \geq 2$, there exists at least one $\td v_i \neq 0$ with $i=2,3,\cdots,n$.
  \item[IV] Delete the \~{} symbol on the newly defined coprime vector $\td {\bf v}$, and repeat Step~II and Step~III. We see that we can reduce $v_1$ by at least one every time we complete Step~II, so for a definite coprime vector ${\bf v}$, we can finally get the simplest coprime vector ${\bf e}=(1,0,\cdots,0)$ within finite number of steps.
\end{enumerate}
\begin{dfn}[Substraction]
The recursion of the previous four steps of transforming an arbitrary coprime vector ${\bf v}$ to the simplest form $\bf{e}$ is called substraction.
\end{dfn}
\noindent It could be seen that the process of substraction is just the exchange of elements of the vector, multiplication of the element with $-1$, and replacing one element by the sum of that element and a multiple of another element. The process of substraction is reversible, i.e. that given a substraction of an coprime ${\bf v}$ to $\bf{e}$, there is a fixed process of transforming $\bf{e}$ to the same ${\bf v}$.
\begin{dfn}[Addition]
The reverse of substraction is called addition.
\end{dfn}

For the simple coprime vector ${\bf e}=(1,0,\cdots,0)$, there are more than one $SL(n,Z)$ matrices that has the vector as the first row. The simple example is the $n\times n$ unit matrix, in fact all the lower triangle matrices with diagonal elements being one are the required matrices, and moreover a unitary elementary operation, or the combination of two anti-unitary elementary operations, of the required matrix with the first row intact also gives the required matrix. So we have a large scope of options.
Suppose that we just choose one of the required matrices $\L^i_{\phantom{i}j}$ for the vector $\bf e$, and then we can make the addition operation of a general coprime vector ${\bf v}$ for all the rows of the matrix at the same time, and effectively we are doing the unitary or anti-unitary elementary column operations for the whole matrix. After the operations the first row of the matrix $\L^i_{\phantom{i}j}$ becomes the elements of vector $v_i$, i.e. that $v_i=\L^1_{\phantom{1}i}$ for $i=1,2,\cdots,n$, and also $\L^i_{\phantom{i}j} \in SL(n,Z)$ or $\L^i_{\phantom{i}j} \in AL(n,Z)$. If $\L^i_{\phantom{i}j} \in AL(n,Z)$, we just make an anti-unitary elementary operation with the first row intact to change the $AL(n,Z)$ matrix to an $SL(n,Z)$ matrix. Now we have one required matrix, and still we can make a unitary elementary operation, or the combination of two anti-unitary elementary operations, of required matrix with the first row intact, and this also gives the required matrix. Thus the Theorem~\ref{z56} we proposed are proved.

\end{appendix}

\providecommand{\href}[2]{#2}\begingroup\raggedright\endgroup

%\nocite{*}

%\bibliographystyle{utphys}   %%非常好，期刊，arXiv超链接
%\bibliographystyle{utcaps}     %%非常好，期刊，arXiv超链接
%\bibliographystyle{jhep}    %%好，arXiv超链接
%\bibliographystyle{kp}     %%好，arXiv超链接

%\bibliography{zbib}

\begin{thebibliography}{10}

\bibitem{Strominger:1996sh}
A.~Strominger and C.~Vafa, ``{Microscopic origin of the Bekenstein-Hawking
  entropy},'' \href{http://dx.doi.org/10.1016/0370-2693(96)00345-0}{{\em
  Phys.Lett.} {\bfseries B379} (1996) 99--104},
\href{http://arxiv.org/abs/hep-th/9601029}{{\ttfamily arXiv:hep-th/9601029
  [hep-th]}}.
%%CITATION = HEP-TH/9601029;%%.

\bibitem{Guica:2008mu}
M.~Guica, T.~Hartman, W.~Song, and A.~Strominger, ``{The Kerr/CFT
  Correspondence},'' \href{http://dx.doi.org/10.1103/PhysRevD.80.124008}{{\em
  Phys.Rev.} {\bfseries D80} (2009) 124008},
\href{http://arxiv.org/abs/0809.4266}{{\ttfamily arXiv:0809.4266 [hep-th]}}.
%%CITATION = ARXIV:0809.4266;%%.

\bibitem{Castro:2010fd}
A.~Castro, A.~Maloney, and A.~Strominger, ``{Hidden Conformal Symmetry of the
  Kerr Black Hole},'' \href{http://dx.doi.org/10.1103/PhysRevD.82.024008}{{\em
  Phys.Rev.} {\bfseries D82} (2010) 024008},
\href{http://arxiv.org/abs/1004.0996}{{\ttfamily arXiv:1004.0996 [hep-th]}}.
%%CITATION = ARXIV:1004.0996;%%.

\bibitem{Bredberg:2011hp}
I.~Bredberg, C.~Keeler, V.~Lysov, and A.~Strominger, ``{Cargese Lectures on the
  Kerr/CFT Correspondence},''
  \href{http://dx.doi.org/10.1016/j.nuclphysbps.2011.04.155}{{\em
  Nucl.Phys.Proc.Suppl.} {\bfseries 216} (2011) 194--210},
\href{http://arxiv.org/abs/1103.2355}{{\ttfamily arXiv:1103.2355 [hep-th]}}.
%%CITATION = ARXIV:1103.2355;%%.

\bibitem{Compere:2012jk}
G.~Compere, ``{The Kerr/CFT correspondence and its extensions: a comprehensive
  review},'' {\em Living Rev.Rel.} {\bfseries 15} (2012) 11,
  \href{http://arxiv.org/abs/1203.3561}{{\ttfamily arXiv:1203.3561 [hep-th]}}.
\url{http://relativity.livingreviews.org/Articles/lrr-2012-11/}.
%%CITATION = ARXIV:1203.3561;%%.

\bibitem{Brown:1986nw}
J.~D. Brown and M.~Henneaux, ``{Central Charges in the Canonical Realization of
  Asymptotic Symmetries: An Example from Three-Dimensional Gravity},''
\href{http://dx.doi.org/10.1007/BF01211590}{{\em Commun.Math.Phys.} {\bfseries
  104} (1986) 207--226}.
%%CITATION = CMPHA,104,207;%%.

\bibitem{Brown:1986ed}
J.~D. Brown and M.~Henneaux, ``{ON THE POISSON BRACKETS OF DIFFERENTIABLE
  GENERATORS IN CLASSICAL FIELD THEORY},''
\href{http://dx.doi.org/10.1063/1.527249}{{\em J.Math.Phys.} {\bfseries 27}
  (1986) 489--491}.
%%CITATION = JMAPA,27,489;%%.

\bibitem{Barnich:2001jy}
G.~Barnich and F.~Brandt, ``{Covariant theory of asymptotic symmetries,
  conservation laws and central charges},''
  \href{http://dx.doi.org/10.1016/S0550-3213(02)00251-1}{{\em Nucl.Phys.}
  {\bfseries B633} (2002) 3--82},
\href{http://arxiv.org/abs/hep-th/0111246}{{\ttfamily arXiv:hep-th/0111246
  [hep-th]}}.
%%CITATION = HEP-TH/0111246;%%.

\bibitem{Barnich:2007bf}
G.~Barnich and G.~Compere, ``{Surface charge algebra in gauge theories and
  thermodynamic integrability},''
  \href{http://dx.doi.org/10.1063/1.2889721}{{\em J.Math.Phys.} {\bfseries 49}
  (2008) 042901},
\href{http://arxiv.org/abs/0708.2378}{{\ttfamily arXiv:0708.2378 [gr-qc]}}.
%%CITATION = ARXIV:0708.2378;%%.

\bibitem{Carlip:2011ax}
S.~Carlip, ``{Extremal and nonextremal Kerr/CFT correspondences},''
  \href{http://dx.doi.org/10.1007/JHEP01(2012)008,
  10.1007/JHEP04(2011)076}{{\em JHEP} {\bfseries 1104} (2011) 076},
\href{http://arxiv.org/abs/1101.5136}{{\ttfamily arXiv:1101.5136 [gr-qc]}}.
%%CITATION = ARXIV:1101.5136;%%.

\bibitem{Carlip:2011vr}
S.~Carlip, ``{Effective Conformal Descriptions of Black Hole Entropy},'' {\em
  Entropy} {\bfseries 13} (2011) 1355--1379,
\href{http://arxiv.org/abs/1107.2678}{{\ttfamily arXiv:1107.2678 [gr-qc]}}.
%%CITATION = ARXIV:1107.2678;%%.

\bibitem{Frolov:1989jh}
V.~P. Frolov and K.~Thorne, ``{Renormalized stress-energy tensor near the
  horizon of a slowly evolving, rotating black hole},''
\href{http://dx.doi.org/10.1103/PhysRevD.39.2125}{{\em Phys.Rev.} {\bfseries
  D39} (1989) 2125--2154}.
%%CITATION = PHRVA,D39,2125;%%.

\bibitem{Chen:2010fr}
B.~Chen, J.~Long, and J.-j. Zhang, ``{Hidden Conformal Symmetry of Extremal
  Black Holes},'' \href{http://dx.doi.org/10.1103/PhysRevD.82.104017}{{\em
  Phys.Rev.} {\bfseries D82} (2010) 104017},
\href{http://arxiv.org/abs/1007.4269}{{\ttfamily arXiv:1007.4269 [hep-th]}}.
%%CITATION = ARXIV:1007.4269;%%.

\bibitem{daCunha:2010jj}
B.~C. da~Cunha and A.~R. de~Queiroz, ``{Kerr-CFT From Black-Hole
  Thermodynamics},'' \href{http://dx.doi.org/10.1007/JHEP08(2010)076}{{\em
  JHEP} {\bfseries 1008} (2010) 076},
\href{http://arxiv.org/abs/1006.0510}{{\ttfamily arXiv:1006.0510 [hep-th]}}.
%%CITATION = ARXIV:1006.0510;%%.

\bibitem{Cvetic:1996kv}
M.~Cvetic and D.~Youm, ``{Entropy of nonextreme charged rotating black holes in
  string theory},'' \href{http://dx.doi.org/10.1103/PhysRevD.54.2612}{{\em
  Phys.Rev.} {\bfseries D54} (1996) 2612--2620},
\href{http://arxiv.org/abs/hep-th/9603147}{{\ttfamily arXiv:hep-th/9603147
  [hep-th]}}.
%%CITATION = HEP-TH/9603147;%%.

\bibitem{Larsen:1997ge}
F.~Larsen, ``{A String model of black hole microstates},''
  \href{http://dx.doi.org/10.1103/PhysRevD.56.1005}{{\em Phys.Rev.} {\bfseries
  D56} (1997) 1005--1008},
\href{http://arxiv.org/abs/hep-th/9702153}{{\ttfamily arXiv:hep-th/9702153
  [hep-th]}}.
%%CITATION = HEP-TH/9702153;%%.

\bibitem{Cvetic:1997uw}
M.~Cvetic and F.~Larsen, ``{General rotating black holes in string theory: Grey
  body factors and event horizons},''
  \href{http://dx.doi.org/10.1103/PhysRevD.56.4994}{{\em Phys.Rev.} {\bfseries
  D56} (1997) 4994--5007},
\href{http://arxiv.org/abs/hep-th/9705192}{{\ttfamily arXiv:hep-th/9705192
  [hep-th]}}.
%%CITATION = HEP-TH/9705192;%%.

\bibitem{Cvetic:1997xv}
M.~Cvetic and F.~Larsen, ``{Grey body factors for rotating black holes in
  four-dimensions},''
  \href{http://dx.doi.org/10.1016/S0550-3213(97)00541-5}{{\em Nucl.Phys.}
  {\bfseries B506} (1997) 107--120},
\href{http://arxiv.org/abs/hep-th/9706071}{{\ttfamily arXiv:hep-th/9706071
  [hep-th]}}.
%%CITATION = HEP-TH/9706071;%%.

\bibitem{Cvetic:2009jn}
M.~Cvetic and F.~Larsen, ``{Greybody Factors and Charges in Kerr/CFT},''
  \href{http://dx.doi.org/10.1088/1126-6708/2009/09/088}{{\em JHEP} {\bfseries
  0909} (2009) 088},
\href{http://arxiv.org/abs/0908.1136}{{\ttfamily arXiv:0908.1136 [hep-th]}}.
%%CITATION = ARXIV:0908.1136;%%.

\bibitem{Cvetic:2010mn}
M.~Cvetic, G.~Gibbons, and C.~Pope, ``{Universal Area Product Formulae for
  Rotating and Charged Black Holes in Four and Higher Dimensions},''
  \href{http://dx.doi.org/10.1103/PhysRevLett.106.121301}{{\em Phys.Rev.Lett.}
  {\bfseries 106} (2011) 121301},
\href{http://arxiv.org/abs/1011.0008}{{\ttfamily arXiv:1011.0008 [hep-th]}}.
%%CITATION = ARXIV:1011.0008;%%.

\bibitem{Castro:2012av}
A.~Castro and M.~J. Rodriguez, ``{Universal properties and the first law of
  black hole inner mechanics},''
  \href{http://dx.doi.org/10.1103/PhysRevD.86.024008}{{\em Phys.Rev.}
  {\bfseries D86} (2012) 024008},
\href{http://arxiv.org/abs/1204.1284}{{\ttfamily arXiv:1204.1284 [hep-th]}}.
%%CITATION = ARXIV:1204.1284;%%.

\bibitem{Detournay:2012ug}
S.~Detournay, ``{Inner Mechanics of 3d Black Holes},''
  \href{http://dx.doi.org/10.1103/PhysRevLett.109.031101}{{\em Phys.Rev.Lett.}
  {\bfseries 109} (2012) 031101},
\href{http://arxiv.org/abs/1204.6088}{{\ttfamily arXiv:1204.6088 [hep-th]}}.
%%CITATION = ARXIV:1204.6088;%%.

\bibitem{Chen:2012mh}
B.~Chen, S.-x. Liu, and J.-j. Zhang, ``{Thermodynamics of Black Hole Horizons
  and Kerr/CFT Correspondence},''
  \href{http://dx.doi.org/10.1007/JHEP11(2012)017}{{\em JHEP} {\bfseries 1211}
  (2012) 017},
\href{http://arxiv.org/abs/1206.2015}{{\ttfamily arXiv:1206.2015 [hep-th]}}.
%%CITATION = ARXIV:1206.2015;%%.

\bibitem{Kastor:1997gt}
D.~Kastor and J.~H. Traschen, ``{A Very effective string model?},''
  \href{http://dx.doi.org/10.1103/PhysRevD.57.4862}{{\em Phys.Rev.} {\bfseries
  D57} (1998) 4862--4869},
\href{http://arxiv.org/abs/hep-th/9707157}{{\ttfamily arXiv:hep-th/9707157
  [hep-th]}}.
%%CITATION = HEP-TH/9707157;%%.

\bibitem{Lu:2008jk}
H.~Lu, J.~Mei, and C.~Pope, ``{Kerr/CFT Correspondence in Diverse
  Dimensions},'' \href{http://dx.doi.org/10.1088/1126-6708/2009/04/054}{{\em
  JHEP} {\bfseries 0904} (2009) 054},
\href{http://arxiv.org/abs/0811.2225}{{\ttfamily arXiv:0811.2225 [hep-th]}}.
%%CITATION = ARXIV:0811.2225;%%.

\bibitem{Hartman:2008pb}
T.~Hartman, K.~Murata, T.~Nishioka, and A.~Strominger, ``{CFT Duals for Extreme
  Black Holes},'' \href{http://dx.doi.org/10.1088/1126-6708/2009/04/019}{{\em
  JHEP} {\bfseries 0904} (2009) 019},
\href{http://arxiv.org/abs/0811.4393}{{\ttfamily arXiv:0811.4393 [hep-th]}}.
%%CITATION = ARXIV:0811.4393;%%.

\bibitem{Azeyanagi:2008kb}
T.~Azeyanagi, N.~Ogawa, and S.~Terashima, ``{Holographic Duals of Kaluza-Klein
  Black Holes},'' \href{http://dx.doi.org/10.1088/1126-6708/2009/04/061}{{\em
  JHEP} {\bfseries 0904} (2009) 061},
\href{http://arxiv.org/abs/0811.4177}{{\ttfamily arXiv:0811.4177 [hep-th]}}.
%%CITATION = ARXIV:0811.4177;%%.

\bibitem{Chow:2008dp}
D.~D. Chow, M.~Cvetic, H.~Lu, and C.~Pope, ``{Extremal Black Hole/CFT
  Correspondence in (Gauged) Supergravities},''
  \href{http://dx.doi.org/10.1103/PhysRevD.79.084018}{{\em Phys.Rev.}
  {\bfseries D79} (2009) 084018},
\href{http://arxiv.org/abs/0812.2918}{{\ttfamily arXiv:0812.2918 [hep-th]}}.
%%CITATION = ARXIV:0812.2918;%%.

\bibitem{Azeyanagi:2008dk}
T.~Azeyanagi, N.~Ogawa, and S.~Terashima, ``{The Kerr/CFT Correspondence and
  String Theory},'' \href{http://dx.doi.org/10.1103/PhysRevD.79.106009}{{\em
  Phys.Rev.} {\bfseries D79} (2009) 106009},
\href{http://arxiv.org/abs/0812.4883}{{\ttfamily arXiv:0812.4883 [hep-th]}}.
%%CITATION = ARXIV:0812.4883;%%.

\bibitem{Chen:2012ps}
B.~Chen and J.-j. Zhang, ``{RN/CFT Correspondence From Thermodynamics},''
  \href{http://arxiv.org/abs/1212.1959}{{\ttfamily arXiv:1212.1959 [hep-th]}}.
to appear in \emph{JHEP}.
%%CITATION = ARXIV:1212.1959;%%.

\bibitem{Chen:2012pt}
B.~Chen and J.-j. Zhang, ``{Electromagnetic Duality in Dyonic RN/CFT
  Correspondence},''
\href{http://arxiv.org/abs/1212.1960}{{\ttfamily arXiv:1212.1960 [hep-th]}}.
%%CITATION = ARXIV:1212.1960;%%.

\bibitem{Krishnan:2010pv}
C.~Krishnan, ``{Hidden Conformal Symmetries of Five-Dimensional Black Holes},''
  \href{http://dx.doi.org/10.1007/JHEP07(2010)039}{{\em JHEP} {\bfseries 1007}
  (2010) 039},
\href{http://arxiv.org/abs/1004.3537}{{\ttfamily arXiv:1004.3537 [hep-th]}}.
%%CITATION = ARXIV:1004.3537;%%.

\bibitem{Chen:2010ywa}
C.-M. Chen, Y.-M. Huang, J.-R. Sun, M.-F. Wu, and S.-J. Zou, ``{Twofold Hidden
  Conformal Symmetries of the Kerr-Newman Black Hole},''
  \href{http://dx.doi.org/10.1103/PhysRevD.82.066004}{{\em Phys.Rev.}
  {\bfseries D82} (2010) 066004},
\href{http://arxiv.org/abs/1006.4097}{{\ttfamily arXiv:1006.4097 [hep-th]}}.
%%CITATION = ARXIV:1006.4097;%%.

\bibitem{Chen:2011wm}
B.~Chen and J.-j. Zhang, ``{Novel CFT Duals for Extreme Black Holes},''
  \href{http://dx.doi.org/10.1016/j.nuclphysb.2011.11.014}{{\em Nucl.Phys.}
  {\bfseries B856} (2012) 449--474},
\href{http://arxiv.org/abs/1106.4148}{{\ttfamily arXiv:1106.4148 [hep-th]}}.
%%CITATION = ARXIV:1106.4148;%%.

\bibitem{Chen:2011kt}
B.~Chen and J.-j. Zhang, ``{General Hidden Conformal Symmetry of 4D Kerr-Newman
  and 5D Kerr Black Holes},''
  \href{http://dx.doi.org/10.1007/JHEP08(2011)114}{{\em JHEP} {\bfseries 1108}
  (2011) 114},
\href{http://arxiv.org/abs/1107.0543}{{\ttfamily arXiv:1107.0543 [hep-th]}}.
%%CITATION = ARXIV:1107.0543;%%.

\bibitem{Strominger:1997eq}
A.~Strominger, ``{Black hole entropy from near horizon microstates},'' {\em
  JHEP} {\bfseries 9802} (1998) 009,
\href{http://arxiv.org/abs/hep-th/9712251}{{\ttfamily arXiv:hep-th/9712251
  [hep-th]}}.
%%CITATION = HEP-TH/9712251;%%.

\bibitem{Wang:2010qv}
Y.-Q. Wang and Y.-X. Liu, ``{Hidden Conformal Symmetry of the Kerr-Newman Black
  Hole},'' \href{http://dx.doi.org/10.1007/JHEP08(2010)087}{{\em JHEP}
  {\bfseries 1008} (2010) 087},
\href{http://arxiv.org/abs/1004.4661}{{\ttfamily arXiv:1004.4661 [hep-th]}}.
%%CITATION = ARXIV:1004.4661;%%.

\bibitem{Chen:2010xu}
B.~Chen and J.~Long, ``{Real-time Correlators and Hidden Conformal Symmetry in
  Kerr/CFT Correspondence},''
  \href{http://dx.doi.org/10.1007/JHEP06(2010)018}{{\em JHEP} {\bfseries 1006}
  (2010) 018},
\href{http://arxiv.org/abs/1004.5039}{{\ttfamily arXiv:1004.5039 [hep-th]}}.
%%CITATION = ARXIV:1004.5039;%%.

\bibitem{Chen:2012yd}
B.~Chen and J.-j. Zhang, ``{Holographic Descriptions of Black Rings},''
  \href{http://dx.doi.org/10.1007/JHEP11(2012)022}{{\em JHEP} {\bfseries 1211}
  (2012) 022},
\href{http://arxiv.org/abs/1208.4413}{{\ttfamily arXiv:1208.4413 [hep-th]}}.
%%CITATION = ARXIV:1208.4413;%%.

\bibitem{Bardeen:1973gs}
J.~M. Bardeen, B.~Carter, and S.~Hawking, ``{The Four laws of black hole
  mechanics},''
\href{http://dx.doi.org/10.1007/BF01645742}{{\em Commun.Math.Phys.} {\bfseries
  31} (1973) 161--170}.
%%CITATION = CMPHA,31,161;%%.

\bibitem{Chen:2010ni}
B.~Chen and C.-S. Chu, ``{Real-Time Correlators in Kerr/CFT Correspondence},''
  \href{http://dx.doi.org/10.1007/JHEP05(2010)004}{{\em JHEP} {\bfseries 1005}
  (2010) 004},
\href{http://arxiv.org/abs/1001.3208}{{\ttfamily arXiv:1001.3208 [hep-th]}}.
%%CITATION = ARXIV:1001.3208;%%.

\bibitem{Cvetic:1996xz}
M.~Cvetic and D.~Youm, ``{General rotating five-dimensional black holes of
  toroidally compactified heterotic string},''
  \href{http://dx.doi.org/10.1016/0550-3213(96)00355-0}{{\em Nucl.Phys.}
  {\bfseries B476} (1996) 118--132},
\href{http://arxiv.org/abs/hep-th/9603100}{{\ttfamily arXiv:hep-th/9603100
  [hep-th]}}.
%%CITATION = HEP-TH/9603100;%%.

\bibitem{Zhang:2013}
J.-j. Zhang, ``{Thermodynamics Method of Black Hole/CFT Correspondence and
  U-duality},''. work in progress.

\bibitem{Banados:1992gq}
M.~Banados, M.~Henneaux, C.~Teitelboim, and J.~Zanelli, ``{Geometry of the
  (2+1) black hole},'' \href{http://dx.doi.org/10.1103/PhysRevD.48.1506}{{\em
  Phys.Rev.} {\bfseries D48} (1993) 1506--1525},
\href{http://arxiv.org/abs/gr-qc/9302012}{{\ttfamily arXiv:gr-qc/9302012
  [gr-qc]}}.
%%CITATION = GR-QC/9302012;%%.

\bibitem{Chen:2010bh}
B.~Chen and J.~Long, ``{On Holographic description of the Kerr-Newman-AdS-dS
  black holes},'' \href{http://dx.doi.org/10.1007/JHEP08(2010)065}{{\em JHEP}
  {\bfseries 1008} (2010) 065},
\href{http://arxiv.org/abs/1006.0157}{{\ttfamily arXiv:1006.0157 [hep-th]}}.
%%CITATION = ARXIV:1006.0157;%%.

\bibitem{Caldarelli:1999xj}
M.~M. Caldarelli, G.~Cognola, and D.~Klemm, ``{Thermodynamics of
  Kerr-Newman-AdS black holes and conformal field theories},''
  \href{http://dx.doi.org/10.1088/0264-9381/17/2/310}{{\em Class.Quant.Grav.}
  {\bfseries 17} (2000) 399--420},
\href{http://arxiv.org/abs/hep-th/9908022}{{\ttfamily arXiv:hep-th/9908022
  [hep-th]}}.
%%CITATION = HEP-TH/9908022;%%.

\bibitem{Gibbons:2004ai}
G.~Gibbons, M.~Perry, and C.~Pope, ``{The First law of thermodynamics for
  Kerr-anti-de Sitter black holes},''
  \href{http://dx.doi.org/10.1088/0264-9381/22/9/002}{{\em Class.Quant.Grav.}
  {\bfseries 22} (2005) 1503--1526},
\href{http://arxiv.org/abs/hep-th/0408217}{{\ttfamily arXiv:hep-th/0408217
  [hep-th]}}.
%%CITATION = HEP-TH/0408217;%%.

\bibitem{Chen:2013aza}
B.~Chen, J.-j. Zhang, J.-d. Zhang, and D.-l. Zhong, ``{Aspects of Warped
  AdS$_3$/CFT$_2$ Correspondence},''
\href{http://arxiv.org/abs/1302.6643}{{\ttfamily arXiv:1302.6643 [hep-th]}}.
%%CITATION = ARXIV:1302.6643;%%.

\end{thebibliography}

\end{document}